\documentclass[12pt]{article}
\usepackage{amsmath}
\usepackage{amsfonts}
\usepackage{graphicx}
\usepackage{equation}
  
\newcommand{\di}{\textrm{d}}
\renewcommand{\Re}{\mathfrak{Re}}
\renewcommand{\Im}{\mathfrak{Im}}

\newcommand{\be}{\begin{equation}}
\newcommand{\ben}{\begin{subequations}}
\newcommand{\een}{\end{subequations}}
\newcommand{\beq}{\begin{eqalignno}}
\newcommand{\eeq}{\end{eqalignno}}
\newcommand{\ee}{\end{equation}}

\newcommand{\intkdn}[6]{C_0(0,0,m_{\tilde{#5}{#6}},
 m_{\tilde{#1}{#2}}, m_{\tilde{#3}{#4}}) }

\newcommand{\intkdz}[6]{C_{D2}(0,0,m_{\tilde{#5}{#6}},
 m_{\tilde{#1}{#2}}, m_{\tilde{#3}{#4}}) }
\newcommand{\intkdd}[6]{C_{D3}(0,0,m_{\tilde{#5}{#6}},
 m_{\tilde{#1}{#2}}, m_{\tilde{#3}{#4}}) }
\newcommand{\intkkd}[6]{C_{\mu\nu}(0,0,m_{\tilde{#1}{#2}},
    m_{\tilde{#3}{#4}},m_{\tilde{#5}{#6}})\delta^{\mu\nu}}
\newcommand{\intkvn}[8]{D_0(0,0,0,m_{\tilde{#7}{#8}},
 m_{\tilde{#1}{#2}}, m_{\tilde{#3}{#4}}, m_{\tilde{#5}{#6}}) }
\newcommand{\intkve}[8]{D_{D1}(0,0,0,m_{\tilde{#7}{#8}},
 m_{\tilde{#1}{#2}}, m_{\tilde{#3}{#4}}, m_{\tilde{#5}{#6}}) }
\newcommand{\intkvz}[8]{D_{D2}(0,0,0,m_{\tilde{#7}{#8}},
 m_{\tilde{#1}{#2}}, m_{\tilde{#3}{#4}}, m_{\tilde{#5}{#6}}) }
\newcommand{\intkkv}[8]{D_{\mu\nu}(0,0,0,m_{\tilde{#1}{#2}},
    m_{\tilde{#3}{#4}},m_{\tilde{#5}{#6}},m_{\tilde{#7}{#8}})\delta^{\mu\nu}}

\def\lsim{\:\raisebox{-0.75ex}{$\stackrel{\textstyle<}{\sim}$}\:}

\DeclareMathOperator{\diag}{diag}
 
\begin{document}
\thispagestyle{empty}
\phantom{xxx}
\vspace{0.3cm}
\begin{flushright}
 MPI-PhT/2003-09 \\
 TUM-HEP-500/03 \\
 hep-ph/0302244 \\
May 2003
\end{flushright}
\vspace{0.4cm}

\begin{center}
 {\Large\bf Complete 1-Loop Calculation of the T-violating 
            $D-$Parameter in Neutron Decay in the MSSM \\ }
\end{center}

\vspace{1cm}
\centerline{\large{\bf Manuel Drees}$^1$ and {\bf Michael Rauch}$^2$} 
\bigskip
\centerline{\it {}$^1$ Physik Dept. TU M\"unchen, D--85748 Garching,
Germany}
\centerline{\it {}$^2$ MPI f. Physik, F\"ohringer Ring 6, 80805
M\"unchen, Germany}

\vspace{1cm}
\centerline{\bf Abstract}

We investigate the violation of time reversal invariance in the decay
of the free neutron in the framework of the Minimal Supersymmetric
Standard Model (MSSM). The coefficient of the triple product of the
neutron spin and the momenta of electron and neutrino, the so called
$D$ parameter, is computed at one loop order including all diagrams.
We find that $D$ is mainly sensitive to the trilinear $A$ coupling in
the squark sector and to the phase of the coefficient $\mu$ which
mixes the two Higgs superfields. The maximal MSSM contribution using
parameters still allowed by experiment is however at $D \approx
10^{-7}$, while QED final state interactions give a value of $D_{\rm
fsi} = - 2.3 \cdot 10^{-5}$. Explicit expressions for all relevant
diagrams are given in an Appendix.


\newpage
\section{Introduction}

The existence of complex parameters in the Lagrangian can lead to a
violation of time reversal symmetry. In the standard model the phase
of the CKM matrix is the only phase which cannot be eliminated by a
field redefinition. In the MSSM there are additional parameters which
in general cannot be made real:
\begin{itemize}
 \item the coefficient of the term bilinear in the Higgs superfields: $\mu$
 \item two of the three gaugino masses $\tilde{m}_i \qquad i=1,2$
 \item terms mixing left-- and right--handed sfermions $A_f$.
\end{itemize}

The violation of (naive) $T$ symmetry can be tested using an
observable which is odd under applying the $T$ symmetry operator. For
example an odd combination of spins and momenta fulfills this
condition. In this paper we investigate the triple product
\begin{equation}
 \frac{\vec{\sigma}_n}{\sigma_n} \cdot (\vec{p}_e \times \vec{p}_{\bar\nu})
\end{equation}
of the spin of the neutron and the momenta of electron and electron
antineutrino in the decay of free neutrons. The coefficient of this
expression in the decay distribution
\begin{equation} \label{decdist}
\frac{\di \Gamma}{\di E_e \di \cos\theta_{e\bar\nu}} = G_e(E_e) \left\{ 1 +
 D \frac{\vec{\sigma}_n}{\sigma_n} \frac{\vec{p}_e \times
    \vec{p}_{\bar\nu}}{E_e E_{\bar\nu}} + \dots \right\}
\end{equation}
where $G_e$ is the tree level expression, is called the $D$ parameter.
It offers a distinct possibility to search for $T$ symmetry violation
in neutron decay, besides the electric dipole moment of the
neutron. An experimental effort is currently underway at the ILL
Grenoble to improve the measurement of, or bound on, $D$. A complete
calculation of $D$ in the MSSM therefore seems timely. To the best of
our knowledge, only the gluino loop contribution to $D$ has previously
been calculated \cite{Christova:1993ju}.

In this paper we follow the conventions of Rosiek \cite{Rosiek:1995kg};
that reference also contains expressions for all relevant Feynman
rules. A brief summary of the (somewhat unusual) notation is given in
Appendix A.

\section{$D-$Parameter}

\subsection{Standard Model}

As remarked earlier, the only $T-$violating parameter in the Standard
Model (SM) is the phase in the Kobayashi--Maskawa matrix. It can lead
to violation of $T$ (or $CP$) symmetry only in processes involving all
three generations of quarks. Therefore it can contribute to the
$D-$parameter only starting at the two--loop level. As a result, the truly
$T$ symmetry violating contribution to $D$ is very small in the SM
\cite{Herczeg:1997se},
\begin{equation} \label{dsm}
 D_{SM} \le 10^{-12}.
\end{equation}

Experimentally a complete time reversal, which would consist of motion
reversal {\em and} exchange of the initial and final states, is
unfortunately not possible in neutron decay. Instead, $D$ is odd under
so--called naive time reversal, where only the directions of all spin
and momentum three--vectors are reversed, without exchanging initial
and final state \cite{soldner:2001}. While genuine $T$ invariance can
only be violated if some parameters in the fundamental Lagrangian
contain nontrivial complex phases, naive time reversal invariance can
be violated whenever the relevant matrix element has a nonvanishing
imaginary part. This difference is significant, since an imaginary
part in the matrix element, a so--called absorptive phase, can also
originate from loop corrections which respect genuine $T$
invariance. In the present case these are due to QED final state
interactions between the proton and the electron. Note that a loop
diagram gives an absorptive phase only if the particles in the loop
can be on--shell; this leads to an additional phase space suppression
factor of order $E_e/m_n$, where $m_n$ is the neutron mass and $E_e$
the energy of the electron in the neutron rest frame. The total
contribution from final state interactions is therefore quite small
\cite{Callan:1967},
\begin{equation} \label{fsi}
\left| D_{\rm fsi} \right| \leq 2.3 \cdot 10^{-5};
\end{equation} 
the bound is saturated at the kinematic maximum of $E_e$. ``New
physics'' contributions to $D$ that are much smaller than this value
will be very difficult to extract even for arbitrarily small
experimental error, since the prediction (\ref{fsi}) has some
theoretical uncertainties, e.g. due to higher order corrections and
proton form factor effects. Finally, the current experimental
sensitivity \cite{Hagiwara:2002pw} is still well below the prediction
(\ref{fsi}), 
\begin{equation} \label{dexp}
D_{\rm exp} = (-0.6 \pm 1.0) \cdot 10^{-3}.
\end{equation}
However, efforts are underway to improve the sensitivity by nearly an
order of magnitude \cite{soldner:2001}.

\subsection{MSSM}

Let us now turn to the calculation of the $D$ parameter in the
MSSM. We have extended the analysis of ref.\cite{Christova:1993ju} by
including all possible diagrams at one loop order. Four different
types of diagrams can contribute to neutron decay:
\begin{itemize}
 \item vertex correction at the $W-$quark vertex;
 \item vertex correction at the $W-$lepton vertex;
 \item vertex corrections where the exchanged $W$ boson is replaced by
 a charged Higgs boson;
 \item box diagrams.
\end{itemize}

The corrections to the $W-$lepton vertex give a contribution to $D$
that is suppressed by a factor $\frac {m_e} {m_p} \simeq 5\cdot
10^{-4}$, so these diagrams can safely be neglected compared to the
corrections to the $W-$quark vertex. The diagrams with Higgs
boson exchange do not contribute at all to the $D-$parameter, since
they do not contain sufficiently many $\gamma$ matrices to give rise
to a spin correlation. We therefore only need to consider corrections
to the $W-$quark vertex as well as box diagrams.

Since we are computing a contribution to the neutron decay
distribution which has nontrivial dependence on final state momenta,
see eq.(\ref{decdist}), we cannot completely neglect external momenta
when evaluating the loop integrals, even though these momenta are much
smaller than the masses of the superparticles in the loop. However,
after introducing Feynman parameters and shifting the loop integration
variable $k$ in such a way that the terms linear in $k$ are eliminated
from the denominator, all terms of order $m_n$ or $m_e$ can be
neglected in the denominator; in other words, external momenta can be
ignored in the loop integrals {\em after} the shift of the loop
momentum. The three-- and four--point functions that appear in our
calculation can therefore easily be reduced to combinations of
two--point functions, as described in Appendix~C.

The coefficients in front of the loop integrals contain three
different kinds of suppression factors. The Dirac algebra can
introduce factors of the nucleon mass, rather than the mass of a
fermionic superparticle. Moreover, certain chargino and neutralino
couplings contain Yukawa couplings to first generation
fermions. Finally, a term may require mixing between $SU(2)$ doublet
and singlet first generation sfermions, which is again proportional to
a first generation Yukawa coupling. Numerically these three
suppression factors are of comparable size, so that a simple counting
of these factors is sufficient to isolate the leading terms.

Let us illustrate these remarks by analyzing the $\tilde \kappa^0 -
\tilde u - \tilde d$ loop correction to the $Wud$ vertex, see
Fig.~10 in Appendix~\ref{feyndiag}. The corresponding
contribution to the $D$ parameter is given in eq.(\ref{eloop1}). We
first note that the three--point function $C_{\mu\nu}$ is ${\cal
O}(1)$, whereas the functions $C_0$, $C_{D2}$ and $C_{D3}$, which are
defined in Appendix~C, are ${\cal O}(1/m_{\rm SUSY}^2)$, where
$m_{SUSY} \sim 0.1$ to 1 TeV is a typical superparticle mass
scale. Let $r \equiv m_n/m_{\rm SUSY}$, and $Y_u$ and $Y_d$ be the $u$
and $d$ quark Yukawa couplings, respectively. From
eqs.(\ref{eloopcoup1}) for the relevant couplings one can then derive
the following behavior for the various terms listed in
eq.(\ref{eloop1}), which we label here by the relevant product of
couplings:
\begin{equation} \label{orders}
A_1 D_1 E_1 \sim {\cal O}(Y_u Y_d); \ \ A_1 C_1 E_1 \sim
{\cal O}(r Y_u); \ \ B_1 D_1 E_1 \sim {\cal O}(r Y_d); \ \ 
B_1 C_1 E_1 \sim {\cal O}(r^2).
\end{equation}
We note that each term has two suppression factors. This is true for
all other loop corrections as well, as can be seen from the results
given in Appendix~B. Terms with an even larger number of suppression
factors have been omitted. 

It should be noted that the terms containing powers of the suppression
factor $r$ have been obtained by replacing the kinematic $u$ and $d$
quark mass by the mass of the proton and neutron, respectively; this
simple approximation goes under the name of ``naive dimensional
analysis'' \cite{diman}. It seems reasonable to take some sort of
long--distance quark mass here, although the use of a constituent
quark mass $\sim m_n/3$ could also be defended. Since the first
generation Yukawa couplings are\footnote{These are short--distance
couplings, to be taken at a momentum scale of order $m_{\rm SUSY}$.}
${\cal O}(10^{-4})$ one might think that $Y_{u,d}$ give a much more
severe suppression than $r$. However, this need not be the case. Only
the imaginary parts of the products of couplings are relevant. Terms
with explicit factors of Yukawa couplings can acquire nontrivial
phases either from gaugino--higgsino mixing in the
chargino--neutralino sector, or from mixing between $SU(2)$ doublet
and singlet sfermions, whereas (in the absence of sflavor mixing)
terms without Yukawa couplings are generally not sensitive to phases
in the sfermion sector. Note also that $Y_d \propto \tan\beta$ for
$\tan\beta \gg 1$. These considerations imply that contributions that
are suppressed by Yukawa couplings are in general not smaller than
those that are suppressed by powers of $r$. Finally, the box diagrams
receive an additional (modest) suppression factor of order
$(m_W/m_{\rm SUSY})^2$. The same factor should also be multiplied to
all terms that do not contain factors of $r$, while terms with only
one power of $r$ receive additional suppression of order $m_W/m_{\rm
SUSY}$. The reason is that heavy superparticles must decouple
quadratically. Such factors result from mixing in the $\tilde \kappa$
sector, and/or from mixing between $SU(2)$ doublet and singlet
squarks. [The latter is also suppressed by the relevant Yukawa
coupling, but this is already included in eq.(\ref{orders}).]
However, for reasonable superparticle masses these additional
suppression factors are much less important than the factors listed in
eq.(\ref{orders}).

\subsection{Restrictions on the parameter space}

In order to make a useful analysis of the $D$ parameter it is
important to know which parts of the parameter space of the MSSM are
still experimentally allowed.  Especially relevant are experiments
which test $T$ symmetry violation in other observables, in particular
the electric dipole moments (EDMs) of electron and neutron, $d_e$ and
$d_n$. Current experimental bounds on these quantities
\cite{Hagiwara:2002pw} impose strong constraints on parameter
space. For superparticle masses of order (a few) hundred GeV, many
combinations of phases are excluded, although in some cases there is a
possibility that one can have small EDMs while retaining large phases
\cite{edmsusy}. For our analysis we have used the formulae for the
EDMs given in \cite{edmsusy}, and checked for each parameter point
that it does not violate the experimental limits
\begin{align} \label{edmlimits}
 \left|d_n\right| \le 0.63\cdot10^{-25} \text{e cm} 
  && \text{(CL=90\%)} \nonumber \\
 -0.005\cdot10^{-26} \text{e cm} \le d_e \le 0.143\cdot10^{-26} \text{e cm}
  && \text{(CL=68\%)} 
\end{align}
Of course, limits on superparticle masses from null results of
experimental searches for these particles at high energy colliders
also have to be respected \cite{Hagiwara:2002pw}.

\section{Numerical Analysis}
\subsection{Choice of Parameters}

We wish to find the maximal supersymmetric contribution to $D$ in the
framework of the $R-$parity conserving MSSM. As well known, the
parameter space of the general MSSM is vast, so some simplifying
assumptions are necessary. In our analysis we have assumed that no
flavor mixing exists in the sfermion sector. This means that the $f
\tilde f \tilde \kappa^0$ and $f \tilde f \tilde g$ vertices are
diagonal in flavor space. One can then easily see from the diagrams
given in Appendix~B that CKM mixing between quarks is not relevant,
i.e. all sfermions in the loop must be of the first generation. On the
other hand, our discussion of eq.(\ref{orders}) showed that it is of
crucial importance to include $L-R$ mixing even for first generation
squarks.

Ignoring all flavor mixing between sfermions may sound like a rather
severe restriction on the parameter space. However, mixing between
first generation sfermions and those of the second or third generation
is strongly constrained by experimental limits on various FCNC
processes \cite{flavmix}. We have checked that including flavor
off--diagonal $LR$ and $RL$ entries of the experimentally allowed
magnitude in the squark mass matrices does not increase the maximal
contribution to $D$ once the constraints (\ref{edmlimits}) on the EDMs
have been imposed. Generally speaking, flavor mixing should not be
very important here, since all external fermions belong to the same
(first) generation.

In the absence of flavor mixing between sfermions our results are
independent of the soft breaking parameters describing the second and
third generation sfermion mass matrices. The same is true for the soft
breaking parameters of the tree--level Higgs potential. We have
initially chosen a rather small value for the ratio of vacuum
expectation values (vevs) $\tan\beta$,
\begin{equation} \label{etb}
 \tan\beta = 3.
\end{equation}
The reason is that supersymmetric contributions to the EDMs increase
with $\tan\beta$ \cite{edmsusy}. Choosing a small value for this
parameter therefore minimizes the impact of the experimental
constraints (\ref{edmlimits}) on the allowed values of the remaining
parameters.

Supersymmetry is a decoupling theory, which means that in the limit
where the masses of the supersymmetric particles go to infinity, the
predictions for all observables approach their SM values. We therefore
consider a relatively light spectrum of superparticles, described by the
following values of the relevant soft breaking parameters:
\ben \label{susypar} \beq
 m_L^2 &= 35\cdot 10^3 \text{ GeV}^2 \qquad
 m_R^2 = 50\cdot 10^3 \text{ GeV}^2 \\
 m_Q^2 &= 150\cdot 10^3 \text{ GeV}^2 \qquad
 m_U^2 = m_D^2 = 200\cdot 10^3 \text{ GeV}^2 \\
 |\mu| &= 450 \text{ GeV} \\
 |m_1| &= 200 \text{ GeV} \qquad
 |m_2| = 400 \text{ GeV} \qquad
 |m_3| = 800 \text{ GeV} 
\eeq \een
Here $m_L$ and $m_R$ are the soft breaking masses for $SU(2)$ doublet
and singlet sleptons, $m_Q$ is the soft breaking mass for $SU(2)$
doublet squarks, and $m_U$ and $m_D$ are the soft breaking masses for
$SU(2)$ singlet squarks with charge $2/3$ and $-1/3$, respectively;
recall that we only need to specify these masses for the first
generation. As mentioned earlier, $\mu$ is the coefficient of the term
coupling the two Higgs superfields in the superpotential. Finally, 
$m_i$ are the soft breaking gaugino masses; the ratios of these masses
are similar to that expected in Grand Unified models with universal
gaugino mass at the unification scale. The choices (\ref{susypar})
lead to a superparticle spectrum that respects all experimental limits
from searches for superparticles, and allows large CP--violating
phases through the cancellation mechanism \cite{edmsusy}. We will
later comment on the effect of lowering the overall SUSY mass scale
from the choice of eqs.(\ref{susypar}).

The remaining parameters are sampled randomly, taking a flat
distribution within specified limits. The trilinear
couplings\footnote{Recall that we are using the convention of
ref.\cite{Rosiek:1995kg}, where the ordinary Yukawa couplings are
explicitly included in the $A-$parameters. In case of first generation
quarks these couplings are roughly of order $10^{-4}$; this explains
the small values of the $A-$parameters in eq.(\ref{apar}).}
\begin{equation} \label{apar}
|A_u| \equiv |A_d| , \, |A_l| \in \left[ 0 , 0.1 \right] \text{ GeV} 
\end{equation}
were chosen so that internal cancellations in the EDMs are possible
but the sfermions do not acquire vevs, which would be the case for too
large values. For simplicity we took the same value for $A_d$ and
$A_u$; $A_l$ is allowed to differ, in order to facilitate independent
cancellations in the supersymmetric contributions to $d_n$ and $d_e$.
Finally the phases of the gaugino masses, $\mu$ and of the trilinear
couplings were varied independently (except $A_d \equiv A_u$) over the
entire possible range,
\begin{equation} \label{phases}
\phi_{m_1}, \, \phi_{m_2}, \, \phi_{A_u} \equiv \phi_{A_d}, \, \phi_{A_l}, \,
\phi_{\mu} \in \left[ 0, 2\pi \right[ .
\end{equation}
Note that by an appropriate field redefinition the phase of the gluino
mass can always be set to zero without loss of generality.

\subsection{Numerical Results}

In this section we display the results of our numerical analysis.
Note that only about 20,000 out of $10^{10}$ tested sets of parameters
satisfied the constraints (\ref{edmlimits}). This illustrates that the
EDMs do indeed severely constrain the allowed combinations of
nontrivial complex phases in the MSSM Lagrangian.

\begin{figure}[ht]
\includegraphics{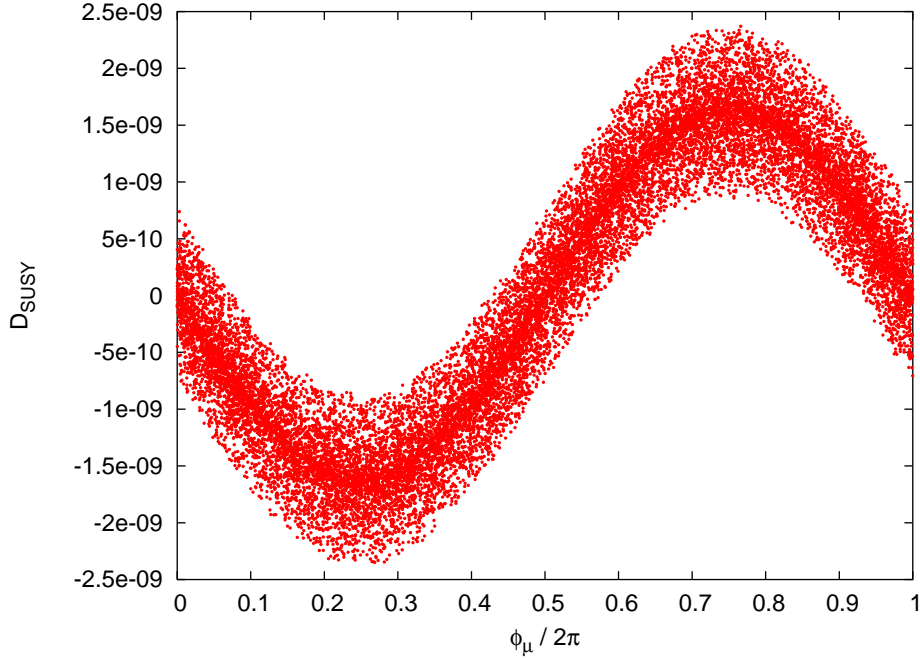}
\caption{Supersymmetric contribution to the $D$ parameter depending on
the phase $\phi_\mu$.}
\label{erg:phimu_D}
\end{figure}

In Fig. \ref{erg:phimu_D} we plot the dependence of the supersymmetric
contribution $D_{\rm SUSY}$ to the $D$ parameter on the phase
$\phi_\mu$. It is easy to see that $D_{\rm SUSY}$ depends strongly on
this phase. There is however also a large variability at fixed
$\phi_\mu$, which shows that $D_{\rm SUSY}$ also depends significantly
on the other parameters.

\begin{figure}
\includegraphics{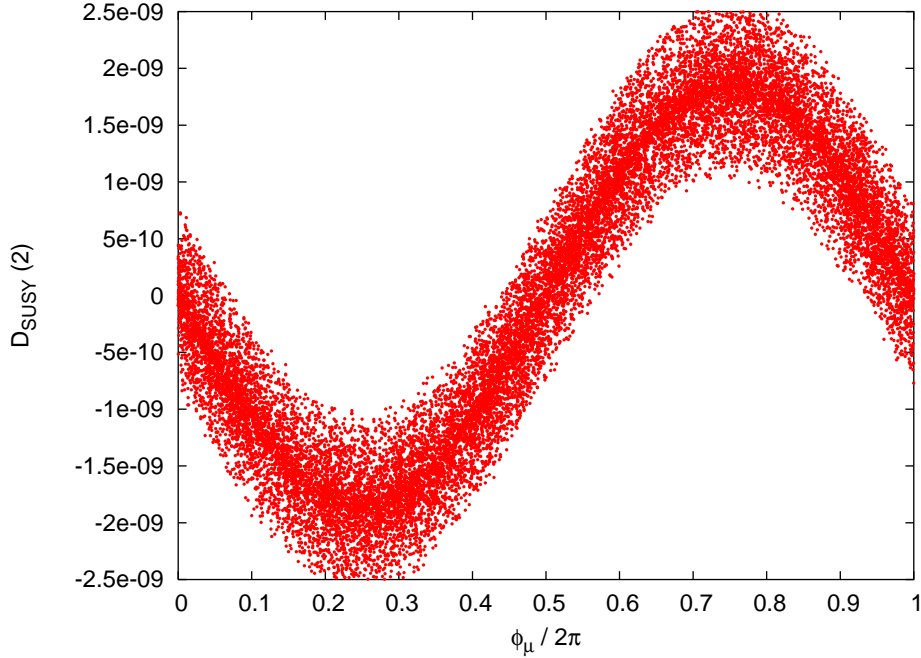}
\caption{The contribution from the leading gluino diagram to $D_{\rm
SUSY}$.}
\label{erg:phimu_D2}
\end{figure}

Before investigating this closer it is useful to identify the diagram
which gives the leading supersymmetric contribution to $D$. Inspection
of the contributing diagrams in Appendix B shows that only one of
them, the quark vertex correction shown in Fig.~11, involves the
strong interactions. One might have thought that the presence of the
gluino in this diagram, which is significantly heavier than the
electroweak gauginos, would partially compensate this
enhancement. However, this would only be true if the gluino was
significantly heavier than the squarks in the loop; such an ordering
of masses is not allowed in the MSSM, since it would lead to tachyonic
squark masses at energy scales just above the gluino mass
\cite{ellwanger}. Indeed we find numerically that the gluino vertex
diagram, which is the only diagram considered in
\cite{Christova:1993ju}, gives the leading contribution to $D_{\rm
SUSY}$. All other diagrams are suppressed by at least one order of
magnitude. This is in spite of the fact that the only phases
contributing to the gluino loop diagram come from squark mixing,
whereas the electroweak loop corrections are also sensitive to phases
from electroweak gaugino--higgsino mixing.\footnote{Partly for this
reason, the chargino and gluino loop contributions to $d_n$ can be of
similar size. Note, however, that in case of the $D$ parameter the
chargino and gluino loop contributions have quite different structure,
i.e. there is no chargino diagram with two squark propagators.} On the
other hand, the relative importance of the other diagrams is increased
by the fact that they {\em all} add destructively to the gluino loop
diagram, i.e. tend to reduce $|D_{\rm SUSY}|$. As a result, the pure
gluino loop contribution to $D_{\rm SUSY}$, shown in
Fig.~\ref{erg:phimu_D2}, is somewhat larger in absolute size than the
total one--loop contribution shown in Fig.~\ref{erg:phimu_D}.

\begin{figure}
\includegraphics[width=0.5\textwidth]{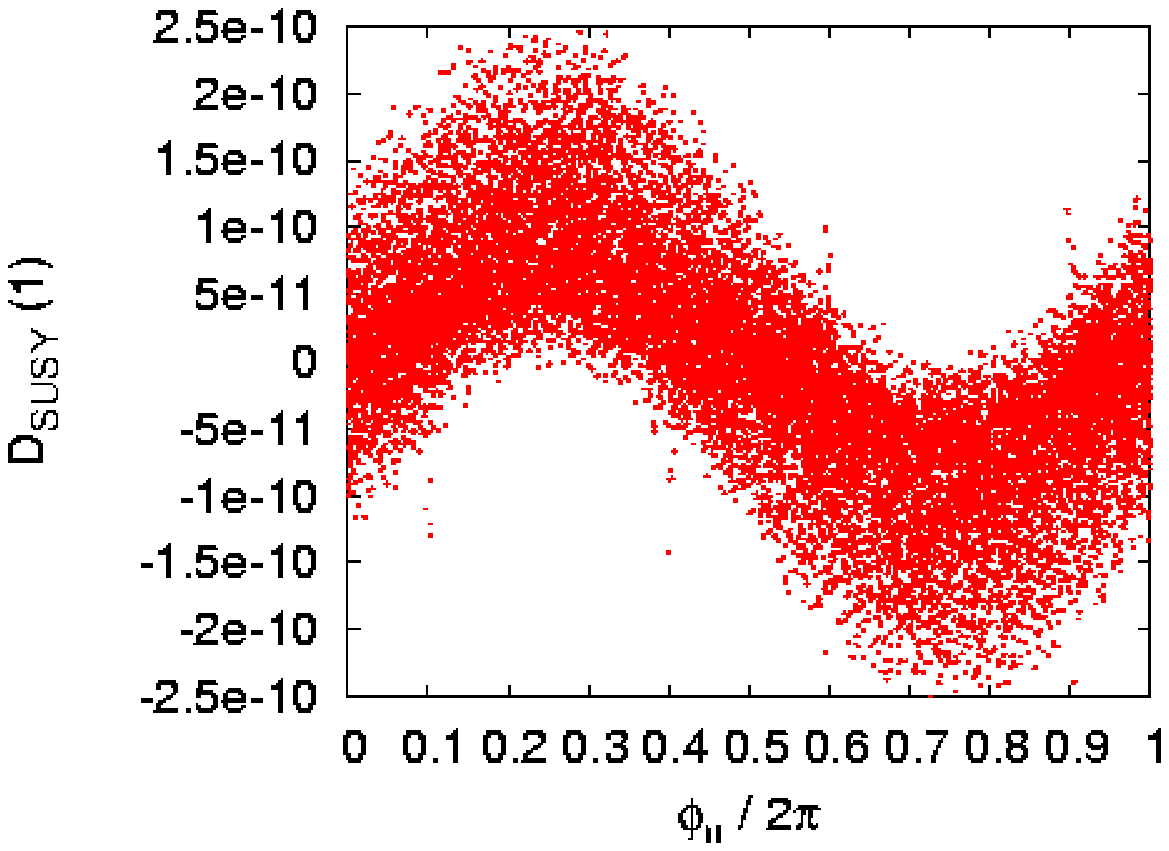}
\includegraphics[width=0.5\textwidth]{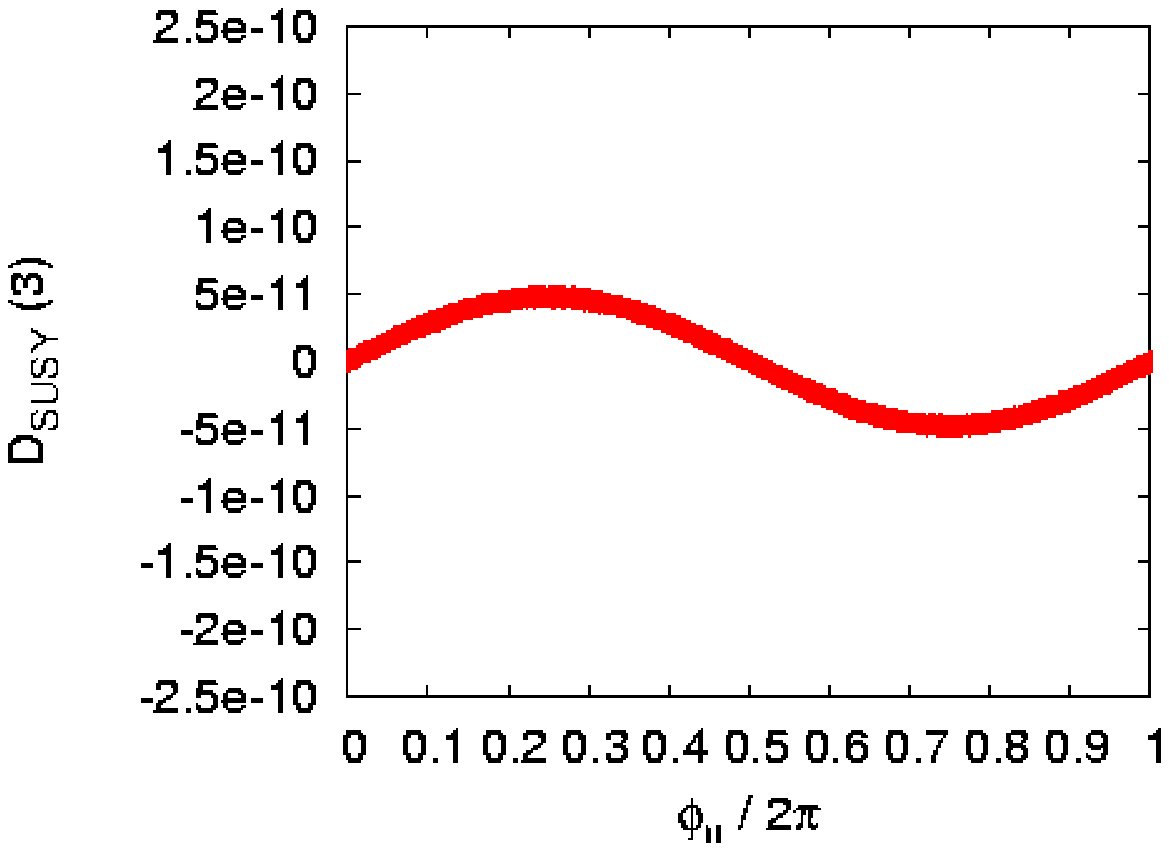}
\includegraphics[width=0.5\textwidth]{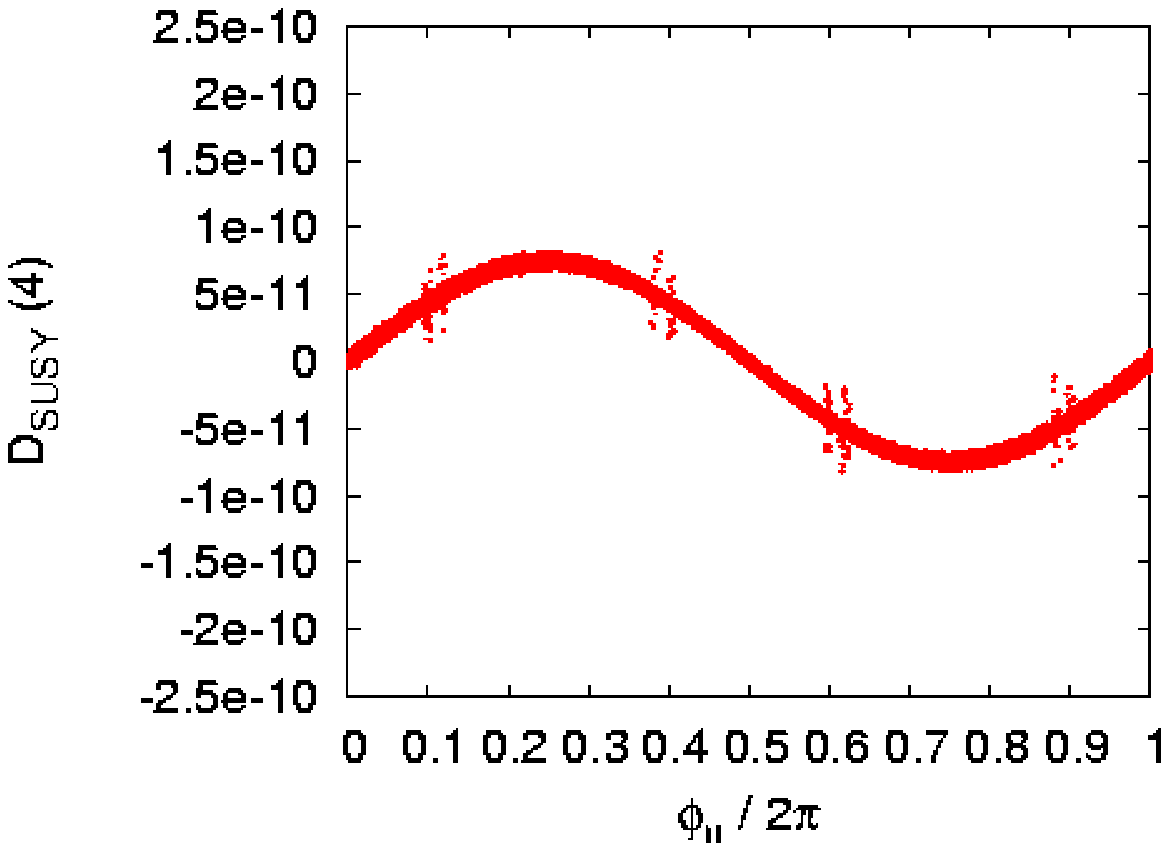}
\includegraphics[width=0.5\textwidth]{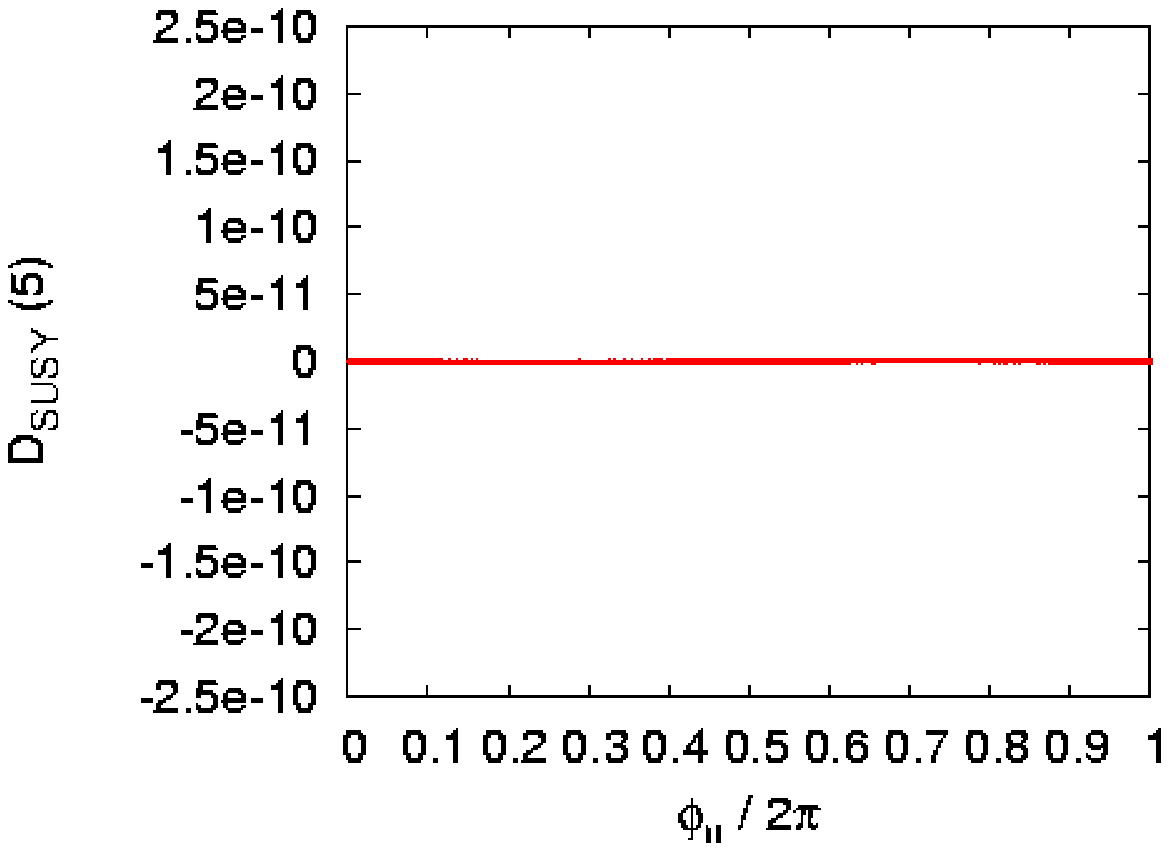}
\begin{center}
\includegraphics[width=0.5\textwidth]{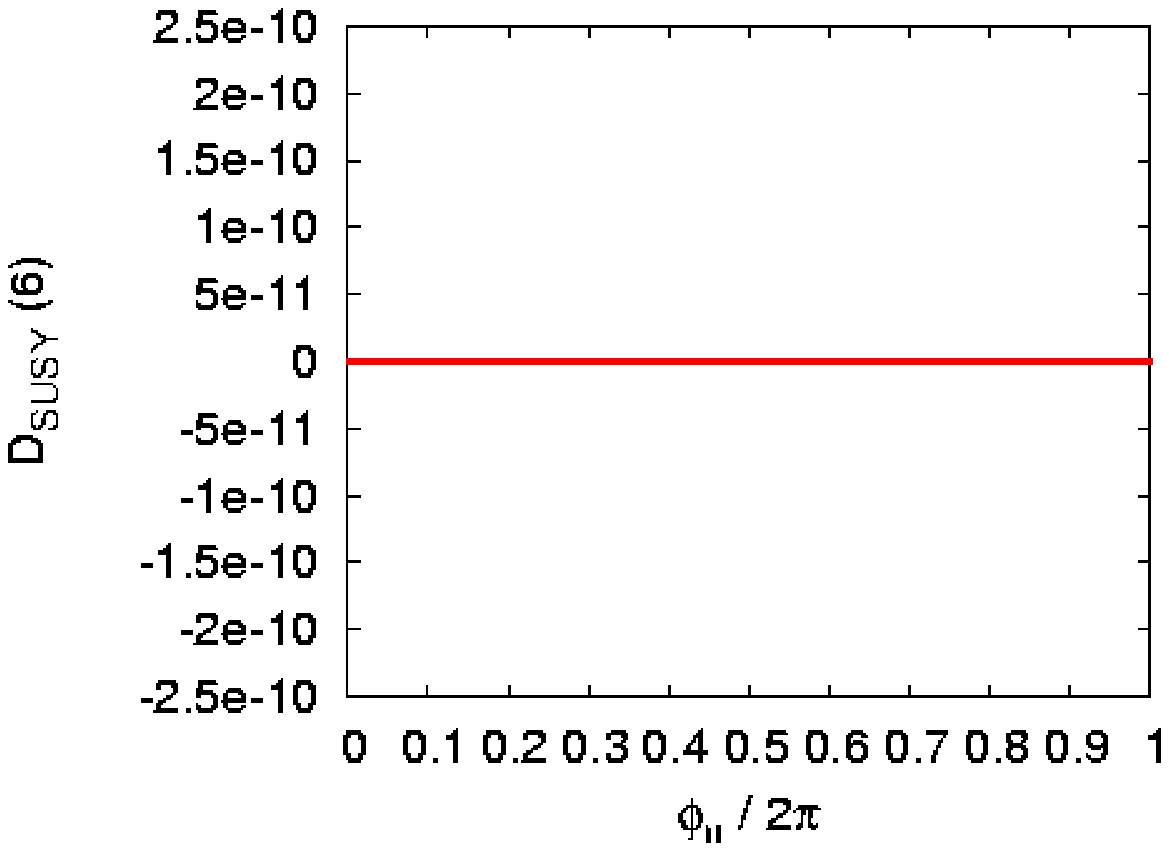}
\end{center}
\caption{Electroweak contributions to $D_{\rm SUSY}$, The numbers in
parentheses correspond to the labeling of diagrams and their
contributions in Appendix B; e.g. $D_{\rm SUSY}(1)$ refers to the
diagram shown in Fig.~10, whose contribution is given by
eq.(\ref{eloop1}).} 
\label{erg:phimu_Drest}
\end{figure}

The contributions of the other five diagrams are shown in
Fig.~\ref{erg:phimu_Drest}. For better readability the $y-$axis was
scaled down by a factor of 10. The numbers in brackets in the label of
the $y$ axis denote the diagram whose contribution is shown in the
corresponding plot. 

We see that the first of the three electroweak vertex corrections,
with a neutralino and two squarks in the loop, gives significantly
larger contributions than diagrams 3 and 4, which have one squark, one
chargino and one neutralino in the loop. This can be understood from
the observation that the $W \tilde \kappa^\pm_i \tilde \kappa^0_j$
vertex only couples two $SU(2)$ gauginos (winos) or two higgsinos
together; this suppresses possible contributions involving the phase
$\phi_1$ associated with the $U(1)_Y$ gaugino (bino). Moreover, the
coupling structure in diagrams 3 and 4 is such that some
gaugino--higgsino mixing is required, whereas diagram 1 gets finite
contributions even without this mixing. This suppresses the
contributions of diagrams 3 and 4 by a factor ${\cal O}(m_W/m_{\rm
SUSY})$ relative to that of diagram 1. Finally, for the given choice
of sparticle masses, the box diagrams 5 and 6 give contributions which
are about two orders of magnitude smaller than that of diagrams 3 and
4. We remarked earlier that the contribution from box diagrams are
suppressed by $(m_W/m_{\rm SUSY})^2 \simeq 1/20$ for our set of
parameters. Moreover, the additional integration over Feynman
parameters required in the $D-$functions appearing in the box
contributions gives another suppression factor $\sim 1/5$ compared to
the $C-$functions appearing in the vertex corrections.

Let us now return to the question which other parameters influence the
size of $D_{\rm SUSY}$. As the gluino diagram is independent of the
phases $\phi_{m_1}$ and $\phi_{m_2}$ these cannot play significant
roles.  If we now restrict $\phi_\mu$ to a small interval around $\pi$
($\phi_\mu =\pi \pm 0.01 \pi$), in accordance with the limits on the
EDMs, it becomes clear that $A_u$ is the second parameter which
determines $D_{\rm SUSY}$ for given sparticle masses. $|A_u|$ largely
determines the amount of $L-R$ mixing between up--type squarks, since
in this case the contribution $\propto |\mu|$ is suppressed by a
factor $\cot\beta$; many of the terms $\propto Y_u$ originate from this
mixing. Since $\mu$ is almost real, $|D_{\rm SUSY}|$ becomes very
small as $|A_u| \rightarrow 0$, as shown in Fig.~4.
Moreover, for (almost) real $\mu$, the phase of $A_u$ ($=A_d$
for our choice of parameters) determines the CP--violating phases in
the squark mixing matrices. This leads us to expect that $|D_{\rm
SUSY}|$ will be maximal if $\phi_{A_u}$ is $\sim \frac{\pi}2$ or $\sim
\frac{3\pi}2$. This is confirmed by Fig.~\ref{erg:phiA_D_rest}. 
($|D_{\rm SUSY}|$ can be small even for these choices of $\phi_{A_u}$
since $|A_u|$ is still varied in Fig.~\ref{erg:phiA_D_rest}; of
course, the value of $\phi_{A_u}$ becomes irrelevant as $|A_u|
\rightarrow 0$.)

\begin{figure}
\includegraphics{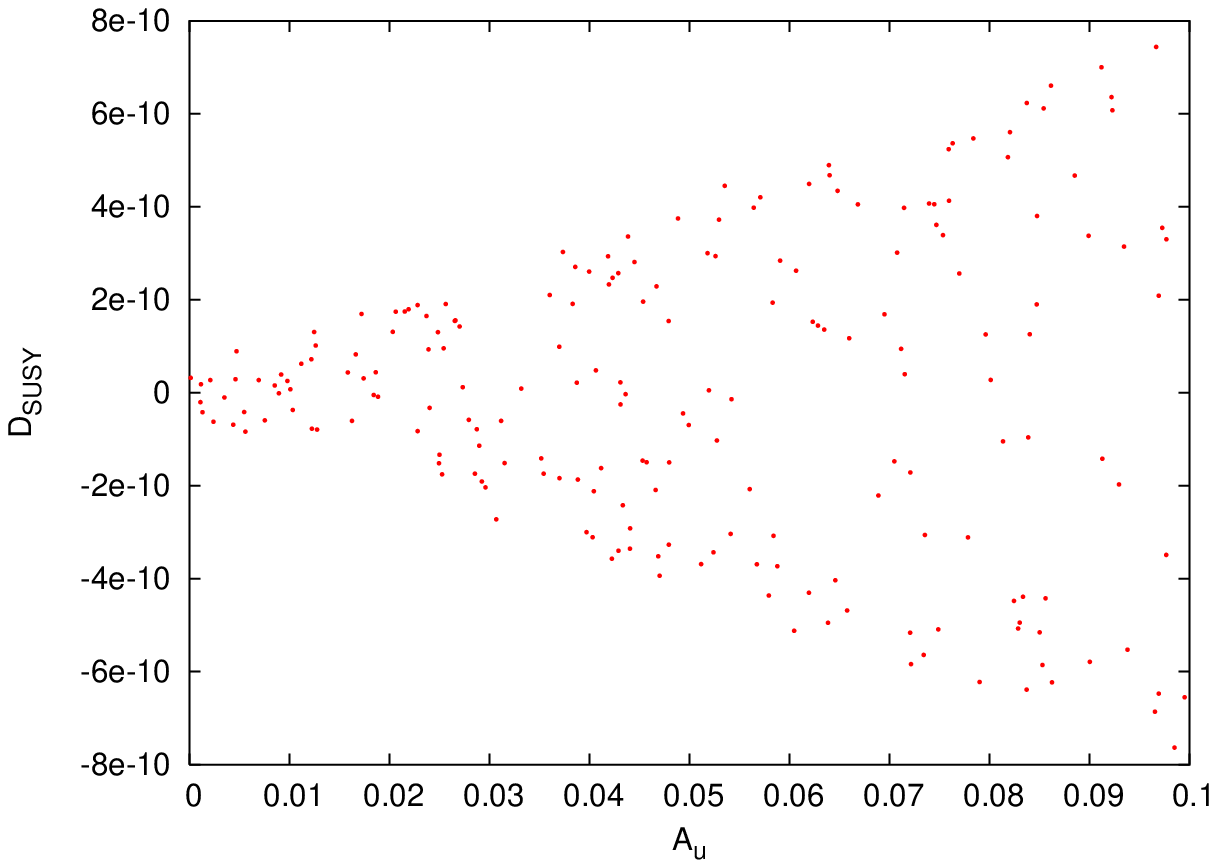}
\caption{$D_{\rm SUSY}$ vs. $A_u$ at $\phi_\mu\approx\pi$}
\label{erg:A_D_rest}
\end{figure}

\begin{figure}
\includegraphics{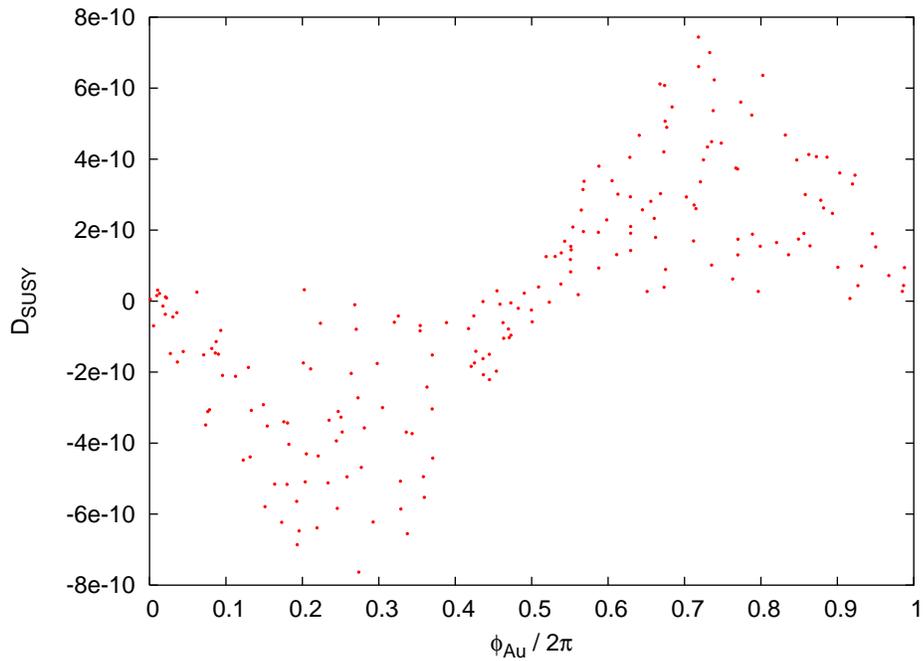}
\caption{$D_{\rm SUSY}$ vs. $\phi_{A_u}$ at $\phi_\mu\approx\pi$}
\label{erg:phiA_D_rest}
\end{figure}

\begin{figure}
\includegraphics{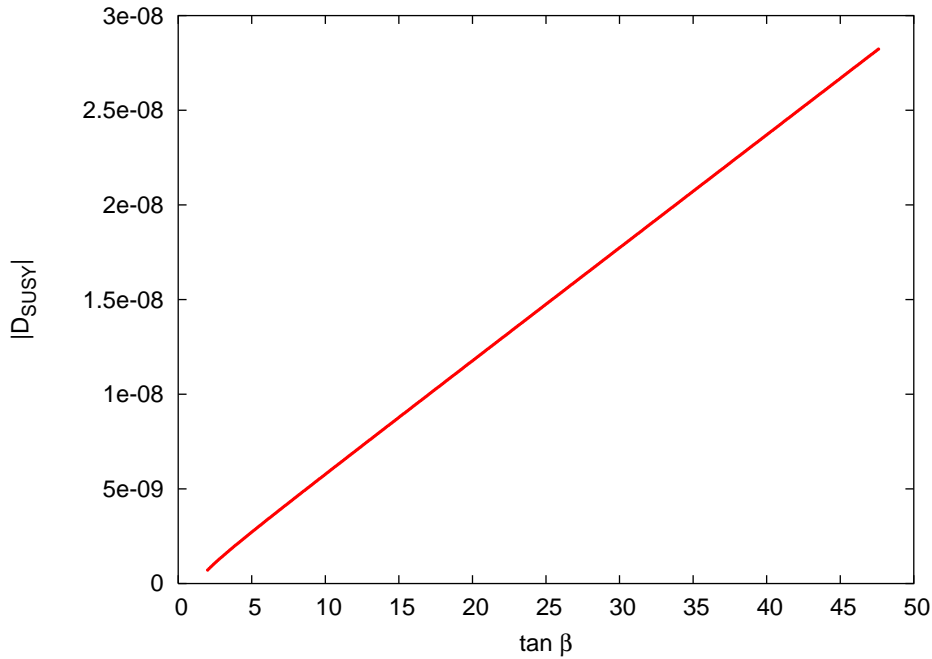}
\caption{$D_{\rm SUSY}$ as function of $\tan\beta$, for the parameter
set (\ref{etanb}).}
\label{erg:tanbeta_D}
\end{figure}

We have also investigated the influence of $\tan\beta$ on $D_{\rm
SUSY}$. To that end the parameter point
\begin{align} \label{etanb}
\phi_{m_2} &= 0.8135\pi & \phi_{m_1} &= 0.09748\pi & \phi_\mu &= 1.4175\pi 
\nonumber \\
|A_u| &= 0.02279 \text{ GeV} & \phi_{A_u} &= 0.2972\pi 
\end{align}
was chosen arbitrarily from the set of points that are allowed for
$\tan\beta = 3$. The $\tan\beta$ dependence of $D_{\rm SUSY}$ is shown
in Fig.~\ref{erg:tanbeta_D}. The almost linear increase results from
the increase of the $d-$quark Yukawa coupling, and hence of $\tilde
d_L - \tilde d_R$ mixing, which is proportional to $1/\cos\beta \simeq
\tan\beta$ for $\tan^2\beta \gg 1$. Here we have neglected the
restrictions from the EDMs; for the set of parameters described by
eqs.(\ref{susypar}), (\ref{apar}) and (\ref{etanb}), these impose the
bound $\tan\beta < 3.5$. On the other hand, since several
contributions to $d_n$ and $d_e$ grow $\propto \tan\beta$,
cancellations can also work at large $\tan\beta$, if some of the other
parameters, e.g. the phases, are changed slightly, without
significantly modifying the prediction for $D_{\rm SUSY}$. However,
since the separate contributions to $d_e$ and $d_n$ grow with
increasing $\tan\beta$, increasingly precise cancellations become
necessary to satisfy the experimental constraints (\ref{edmlimits}).

Both of these observations are confirmed by Fig.~7, which shows
$D_{\rm SUSY}$ for the same parameters as in Fig.~1, except that now
$\tan\beta = 30$. We see that the overall scale of $D_{\rm SUSY}$ is
one order of magnitude larger than in Fig.~1, as expected from the
linear growth shown in Fig.~6. The small number of surviving points
illustrates the difficulty of getting both $|d_e|$ and $|d_n|$
sufficiently small through delicate cancellations. In particular, the
$d_e$ constraint now excludes some region of $\phi_\mu$
altogether.\footnote{Recall that $\phi_\mu$ measures the relative
phase between $\mu$ and the gluino mass $m_3$. For the choice
(\ref{susypar}) of dimensional parameters the $d_e$ constraint limits
the relative phase between $\mu$ and the $SU(2)$ gaugino mass $m_2$ to
narrow bands around 0 and $\pi$ even for small $\tan\beta$, but this
is not visible after scanning over $\phi_{m_2}$.} Recall, however,
that the (s)leptonic corrections to $D_{\rm SUSY}$ are suppressed by a
factor $m_e/m_p$, and are thus negligible. We can therefore vary the
sleptonic soft breaking masses $m_L, \, m_R$ without significantly
changing the prediction for $D_{\rm SUSY}$. In this case the entire
range of $\phi_\mu$ becomes allowed again. We therefore conclude that
the maximal allowed $|D_{\rm SUSY}|$ increases essentially linearly
with $\tan\beta$. However, requiring the bottom Yukawa coupling to be
less than that of the top quark, or at least to be sufficiently small
to not have a Landau pole below the scale of Grand Unification, leads
to the upper bound $\tan\beta \lsim 60$. If the dimensionful
parameters in the squark, gaugino and higgsino sectors are as in
eqs.(9), the maximal value of $|D_{\rm SUSY}|$ is therefore around $5
\cdot 10^{-8}$.

\begin{figure}
\includegraphics{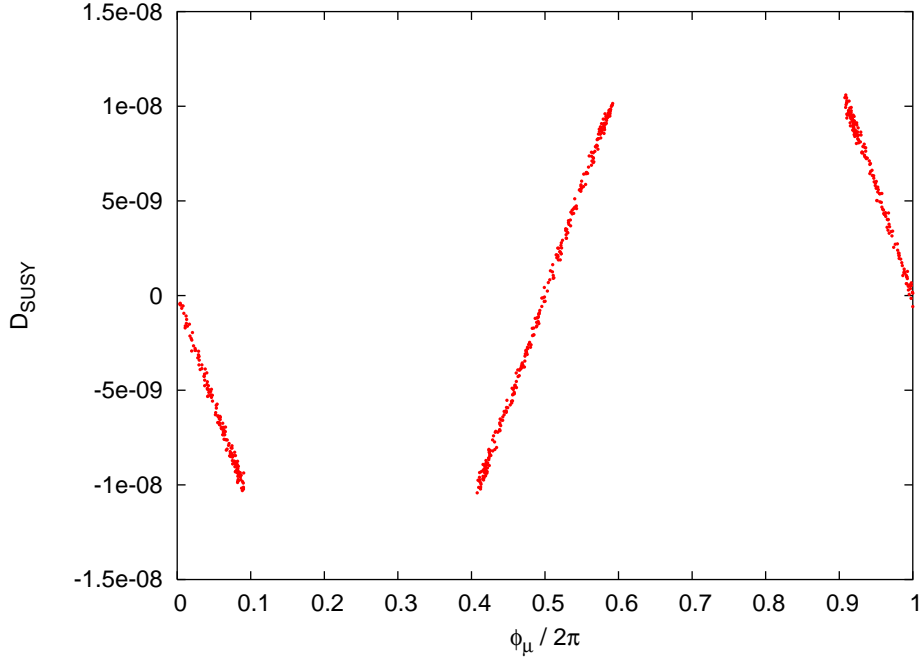}
\caption{$D_{\rm SUSY}$ vs. $\phi_\mu$ for $\tan\beta = 30$.}
\label{erg:phimu_D_ltb}
\end{figure}

Our choice (\ref{susypar}) for the relevant soft breaking masses means
that first generation squarks as well as most charginos and
neutralinos have masses around 400 to 500 GeV. Direct searches for
sparticles allow us to lower these masses by a factor of two to
three. We expect that for fixed phases $D$ scales quadratically with
the overall superparticle mass scale, $D \propto 1/m^2_{\rm SUSY}$. 

This is borne out by Fig.~8, which shows $D_{\rm SUSY}$ for $\tan\beta
= 3$; the other parameters are as in eqs.(\ref{susypar}) and
(\ref{etanb}), except that all quantities with mass dimension are
multiplied with the dimensionless factor $c$, which is varied between
0.4 and 2. Similar to the case of Fig.~6, values of $c$ significantly
different from unity are disallowed by the constraints on $d_e$ and/or
$d_n$, but again this can be fixed by small variations of the
phases. Since $D_{\rm SUSY}$ and the dipole moments show the same
$c^{-2}$ dependence on $c$, increasingly delicate cancellations are
required to satisfy the constraints (\ref{edmlimits}) as $c$ is
reduced. The lower bound $c \geq 0.44$ is in our case set by the lower
bound on the mass of the lightest selectron, $m_{\tilde e} \geq 95$
GeV, from LEP searches \cite{Hagiwara:2002pw}.

\begin{figure}
\includegraphics{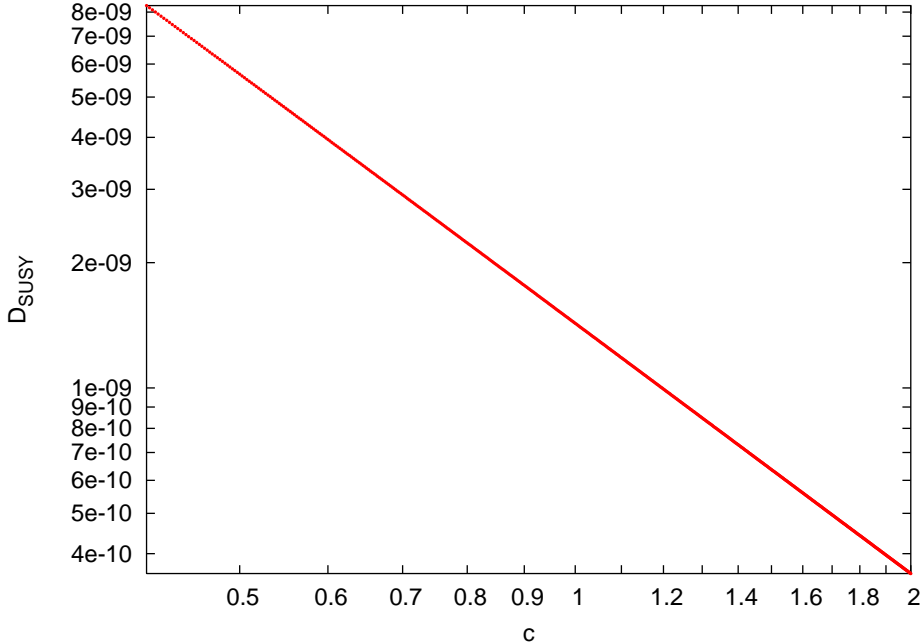}
\caption{$D_{\rm SUSY}$ for $\tan\beta = 3$, where the other
parameters are as in eqs.(\ref{susypar}) and (\ref{etanb}), except
that all mass parameters have been multiplied with the dimensionless
scaling factor $c$.}
\label{erg:phimu_c}
\end{figure}

By simultaneously reducing the overall SUSY mass scale {\em and}
increasing $\tan\beta$ one might therefore in principle be able to
reach values of $|D_{\rm SUSY}|$ slightly above $10^{-7}$. However,
the finetuning required to satisfy the limits on both $d_e$ and $d_n$
then becomes very severe indeed; less than one parameter set in 
$10^9$ will survive.\footnote{Since $d_e$ and $d_n$ have to be
finetuned independently, the overall severity of finetuning scales
like $c^{-4} \cdot \tan^2 \beta.$} Moreover, one would have to choose
soft breaking masses for $\tilde \tau$ sleptons and $\tilde b$ squarks
that are much larger than those for the corresponding first generation
sfermions. Otherwise $L-R$ mixing, which grows $\propto |\mu|
\tan\beta$, would make the lighter $\tilde \tau$ and $\tilde b$ mass
eigenstates much too light, or even tachyonic. Recall that even
$|D_{\rm SUSY}| \simeq 10^{-7}$ is still two orders of magnitude below
the contribution (\ref{fsi}) from final state interactions.

Finally, we have checked for a few cases that changing the ratios of
soft breaking parameters from the choice of eqs.(\ref{susypar}) does
not increase the maximal allowed value of $D_{\rm SUSY}$
significantly. This is not surprising, since the overall scale of
$D_{\rm SUSY}$ is set by the {\em heaviest} superparticle that occurs
in a given loop diagram, whereas the lower bound on the overall mass
scale is essentially set by the {\em lightest} (charged)
superparticle. $|D_{\rm SUSY}|$ will therefore be maximal if the
parameters are chosen such that the mass splitting between
superparticles is relatively small, which is true for the parameters
of eqs.(\ref{susypar}).

\section{Conclusions and Outlook}

In this paper we have analyzed $T$ symmetry violation in the beta decay
of free neutrons via the $D$ parameter. We have extended the analysis of
\cite{Christova:1993ju} by computing all diagrams that occur at one loop
order. We have performed a full scan of the allowed phases and the
magnitude of the $A$ parameters describing mixing between $SU(2)$
singlet and doublet squarks, subject only to the experimental
constraints on the electric dipole moments of the electron and, in
particular, the neutron.

We find that the gluino loop correction to the $W u d$ vertex indeed
gives the leading supersymmetric contribution to $D$. The phase of
$\mu$, which has been neglected in ref.\cite{Christova:1993ju},
crucially influences the size of $D_{\rm SUSY}$. Moreover, allowing
for cancellations between various contributions to the EDMs permits
larger values for the relevant phases, which increases $|D_{\rm
SUSY}|$. We nevertheless find that the maximal contribution to $|D|$
from the $R-$parity conserving MSSM is at least two orders of
magnitude smaller than the contribution (\ref{fsi}) from
electromagnetic final state interactions. This means that even greatly
improved experimental upper bounds on $|D|$ will not lead to new
constraints on this model. On the other hand, a measurement of $D$
which differs significantly from the prediction (\ref{fsi}) would rule
out the $R-$parity conserving MSSM along with the SM. Larger
contributions might be possible in the $R-$parity violating version of
the MSSM. Note that $R-$parity violation only through trilinear terms
in the superpotential does not contribute to the EDMs at one--loop
level \cite{edmrpv}, whereas e.g. baryon--number violating
($\lambda''$) couplings can contribute to $D$ at the one--loop level.

Analogous triple products can also be defined for decays involving
heavier particles than the neutron. Decays of heavier baryons,
e.g. $\Lambda$ or $\Lambda_b$, can be treated using the expressions
given in Appendix~B, since here the external momenta are still much
smaller than $m_{\rm SUSY}$. In these cases the supersymmetric
contributions are expected to be larger by several orders of magnitude
than in case of neutron decay, since $Y_d$ would be replaced by $Y_s$
or even $Y_b$; in addition, the constraints on CP--violating phases of
soft breaking parameters in the second and third generation are much
weaker than for the first generation. Experimental measurements will
probably be difficult in these cases, however.

Even larger supersymmetric contributions can be expected for the
analogous asymmetry in top quark decay. Since the top Yukawa coupling
is ${\cal O}(1)$, electroweak corrections might well be comparable to
SUSY QCD corrections \cite{Christova:1993ju} in this case. Moreover,
the mass ratio $m_t/m_{\rm SUSY}$ is also ${\cal O}(1)$. This again
increases the level of the expected corrections; it also means,
however, that the expansion for small external momenta used in our
calculation is no longer adequate. Note also that the spin of the top
quark cannot be measured directly; nevertheless the $D-$parameter in
top decays does contribute to measurable $T-$odd asymmetries
\cite{top}. The final state interactions for such decays have already
been found to have approximately the same size as for neutron decay
\cite{Liu:1993ne}. A full calculation of supersymmetric contributions
to CP--violation in top decay, including electroweak contributions,
might therefore prove rewarding.

\subsubsection*{Acknowledgements}
We thank Janusz Rosiek for providing us with FORTRAN code for the
diagonalization of the mass matrices and for the evaluation of some
loop functions, and Athanasios Dedes for useful discussions. This work
was partially supported by the SFB 375 of the Deutsche
Forschungsgemeinschaft.

\begin{appendix}

\section{Notation}
\renewcommand{\theequation}{A.\arabic{equation}}
\setcounter{equation}{0}

For the benefit of the reader we have summarized in this appendix the
conventions used in this paper. A complete description including all
expressions for the Feynman rules can be found in
\cite{Rosiek:1995kg}.

There are two charginos $\kappa^\pm_i, i=1,2$ whose mass matrix is
diagonalized by two unitary matrices $Z_+$ and $Z_-$:
\be
 (Z_-)^T \begin{pmatrix}
 m_2 & \frac{e v_2}{\sqrt{2} s_W} \\
 \frac{e v_1}{\sqrt{2} s_W} & \mu
 \end{pmatrix} Z_+ = \diag(m_{\kappa^\pm_1},m_{\kappa^\pm_2})
\ee
This equation does not specify the two matrices $Z_+$ and $Z_-$
uniquely. This can be used to choose both masses positive and sorted
in ascending order.

Similarly, the neutralinos are denoted by $\kappa_i^0, i=1\dots4$ and
the neutralino mass matrix is diagonalized by a unitary matrix $Z_N$
such that
\be
 (Z_N)^T \begin{pmatrix}
  m_1 & 0 & \frac{-e v_1}{2 c_W} & \frac{e v_2}{2 c_W} \\
  0 & m_2 & \frac{e v_1}{2 s_W} & \frac{-e v_2}{2 s_W} \\
  \frac{-e v_1}{2 c_W} & \frac{e v_1}{2 s_W} & 0 & -\mu \\
  \frac{e v_2}{2 c_W} & \frac{-e v_2}{2 s_W} & -\mu & 0 
 \end{pmatrix} Z_N = \diag(m_{\kappa_1^0},\dots,m_{\kappa_4^0})
\ee

As noted in the text, the mass parameter of the $SU(3)$ gauginos can
be taken as real. Therefore the eight gluinos all have a mass of
$m_3$.

Finally, the sfermion mass matrix can be written compactly for all
four different types of sfermions as
\be
 \label{sfermionmixing}
  {\cal M}_{\tilde f}^2 = \begin{pmatrix}
   {m_{{\tilde f}_L}^2}^T + m_f^2 + 
    \frac{e^2\left(v_1^2-v_2^2\right)\left(T_f^3-Q_f s_W^2 \right)}%
     {4 s_W^2 c_W^2} &
   - m_f \left( \kappa \mu^* + \frac{A_f}{Y} \right) \\
   - m_f \left( \kappa \mu + \frac{A_f^*}{Y} \right) &
   m_{{\tilde f}_R}^2 + m_f^2 
     + Q_f \frac{e^2\left(v_1^2-v_2^2\right)}{4 c_W^2}
  \end{pmatrix}
\ee
Here, $\kappa = \cot\beta $ for up--type squarks and $\kappa =
\tan\beta $ for down--type squarks and charged sleptons. $m_{{\tilde
f}_L}$ and $m_{{\tilde f}_R}$ denote the mass parameters and $A_f$ the
coefficient of the trilinear terms from the soft SUSY-breaking terms
in the Lagrangian, and $Y$ is the respective Yukawa coupling. $Q_f$ is
the electromagnetic charge and $T_f^3$ the quantum number of the third
component of the isospin operator. As there is no $SU(2)$ singlet
sneutrino only the upper left element in the matrix of
eq. \ref{sfermionmixing} occurs for the sneutrinos.

These mass matrices can be diagonalized with a unitary matrix each,
yielding
\begin{align} 
 Z_\nu^T {\cal M}_{\tilde\nu}^2 Z_\nu^* 
  &= \diag\left(m_{{\tilde\nu}_1}^2, \dots, m_{{\tilde\nu}_3}^2 \right)&
 Z_L^\dagger {\cal M}_{\tilde e}^2 Z_L 
  &= \diag\left(m_{{\tilde e}_1}^2, \dots, m_{{\tilde e}_6}^2 \right)
   \nonumber\\
 Z_U^T {\cal M}_{\tilde u}^2 Z_U^*
  &= \diag\left(m_{{\tilde u}_1}^2, \dots, m_{{\tilde u}_6}^2 \right)&
 Z_D^\dagger {\cal M}_{\tilde d}^2 Z_D 
  &= \diag\left(m_{{\tilde d}_1}^2, \dots, m_{{\tilde d}_6}^2 \right)
\end{align}

\section{Feynman diagrams}
\label{feyndiag}
\renewcommand{\theequation}{B.\arabic{equation}}
\setcounter{equation}{0}

Altogether six Feynman diagrams were computed. The vertex corrections at
the W-lepton vertex are, as already mentioned, suppressed by a factor of
$\frac{m_e}{m_p}\simeq 5\cdot 10^{-4}$ and can therefore be neglected.
We first give the tree level expression for the differential decay
distribution, since it enters the definition of the
$D-$parameter: 
\be \label{ddef}
D = \frac{\di \Gamma_i}{\di E_e \di\cos\theta_{e\bar\nu}} \ \slash\  
    \left( \frac{\di \Gamma_{\text{tree}}}{\di E_e \di\cos\theta_{e\bar\nu}}
           \cdot \vec{s_n} \cdot \left(
           \frac{\vec{p}_e \times \vec{p}_{\bar\nu}}{E_e E_{\bar\nu}} \right)
    \right) 
\ee
All necessary traces have been computed with the help of FORM
\cite{form}; in many cases the results have been checked using manual
calculations.

\paragraph{Tree level result}

\begin{figure}[ht]
\begin{center}
 \includegraphics{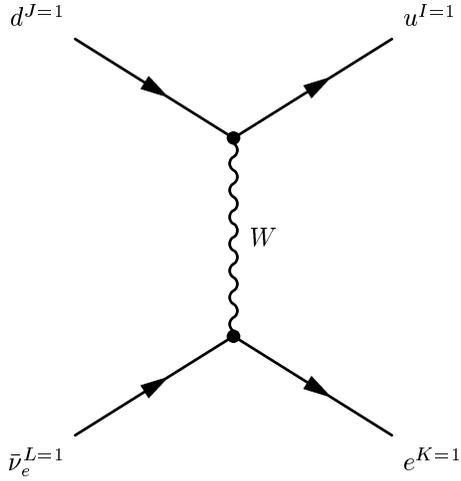}
 \caption{Tree level diagram describing neutron decay.}
 \label{formeln:feyn_tree}
\end{center}
\end{figure}

\begin{align}
\frac{\di \Gamma_{\text{tree}}}{\di E_e \di\cos\theta_{e\bar\nu}} =&
\frac{E_{\bar\nu}^2\sqrt{E_e^2-m_e^2}}
  {8 \pi^3 m_W^4 \left(m_n-E_e-E_{\bar\nu}\right)} E_e m_n U^2 V^2
\end{align}
with
\begin{align}
U &= - \frac{e}{\sqrt{2}\sin\theta_W} C^{IJ} \\
V &= - \frac{e}{\sqrt{2}\sin\theta_W} \delta^{KL}.
\end{align}
Here $e$ is the QED coupling constant, $\theta_W$ the weak mixing
angle, and $C$ the quark flavor mixing matrix. In our numerical
calculation we have ignored all flavor mixing.

\paragraph{Diagram 1: Neutralino-$\tilde u$-$\tilde d$ vertex correction}~\\

\begin{figure}[ht] \label{loop1}
\begin{center}
 \includegraphics{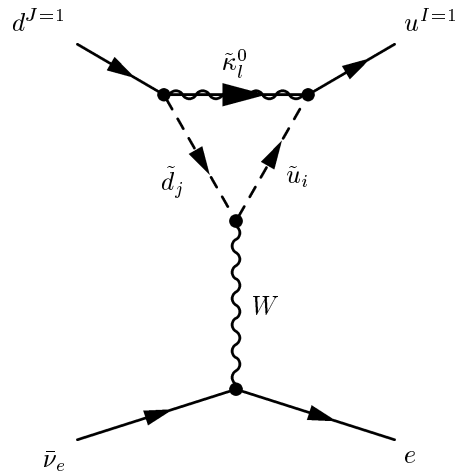}
 \caption{Vertex correction with neutralino-$\tilde u$-$\tilde d$ loop.}
\end{center}
\end{figure}

\beq \label{eloop1}
\frac{\di \Gamma_1}{\di E_e \di\cos\theta_{e\bar\nu}} 
=& \frac{\di \Gamma_{\text{tree}}}{\di E_e \di\cos\theta_{e\bar\nu}}
    \cdot \vec{s_n} \cdot \left( 
    \frac{\vec{p}_e \times \vec{p}_{\bar\nu}}{E_e E_{\bar\nu}} \right) 
    \frac1{4 m_n U^2 V^2} 
   \frac{ \Re{\left(UV^2\right)} } {16 \pi^2} \nonumber \\ 
&\Bigg[ \Bigl( -2 \Im\left(A_1 D_1 E_1\right) m_p \Bigr) 
\intkkd{u}{_i}{d}{_j}{\kappa}{^0_l} \\
& + \Bigl( 4 \Im\left(A_1 C_1 E_1\right) 
  + 4 \Im\left(B_1 D_1 E_1\right) \Bigr) m_{\tilde{\kappa}^0_l} m_n^2 
\intkdz{u}{_i}{d}{_j}{\kappa}{^0_l}
\nonumber \\& 
+ 4 \Im\left(B_1 C_1 E_1\right) m_n^3 
\left( \intkdz{u}{_i}{d}{_j}{\kappa}{^0_l} 
       - \intkdd{u}{_i}{d}{_j}{\kappa}{^0_l} \right)
\Bigg] \nonumber \\
& + \dots \nonumber
\eeq
with
\ben \label{eloopcoup1} \beq
A_1 =& \frac{2\sqrt2e}{3\cos\theta_W}Z_U^{\left(I+3\right)i}Z_N^{1l}
        - Y_u^I Z_N^{4l} Z_U^{Ii} \\
B_1 =& - \frac{e}{\sqrt2\sin\theta_W\cos\theta_W} Z_U^{Ii} \left( \frac13
        Z_N^{1l*} \sin\theta_W + Z_N^{2l*} \cos\theta_W \right)
- Y_u^I Z_N^{4l*}Z_U^{\left(I+3\right)i} \\
C_1 =& - \frac{e}{\sqrt2\sin\theta_W\cos\theta_W} Z_D^{Jj} \left( \frac13
        Z_N^{1l} \sin\theta_W - Z_N^{2l} \cos\theta_W \right)
+ Y_d^J Z_D^{\left(J+3\right)j} Z_N^{3l} \\
D_1 =& \frac{-\sqrt2e}{3\cos\theta_W} Z_D^{\left(J+3\right)j} Z_N^{1l*}
        +Y_d^J Z_D^{Jj} Z_N^{3l*} \\
E_1 =& - \frac{e}{\sqrt{2}\sin\theta_W} Z_D^{Jj*} Z_U^{Ii*} C^{IJ} 
\eeq \een
The dots $(\dots)$ denote additional terms in the decay distribution
that do not contribute to $D$. Eqs. (\ref{eloop1}) and
(\ref{eloopcoup1}), as well as all subsequent expressions, are given
in the notation of Rosiek \cite{Rosiek:1995kg}.

\paragraph{Diagram 2: Gluino-$\tilde u$-$\tilde d$ vertex correction.}~\\

\begin{figure}[ht]
\begin{center}
 \includegraphics{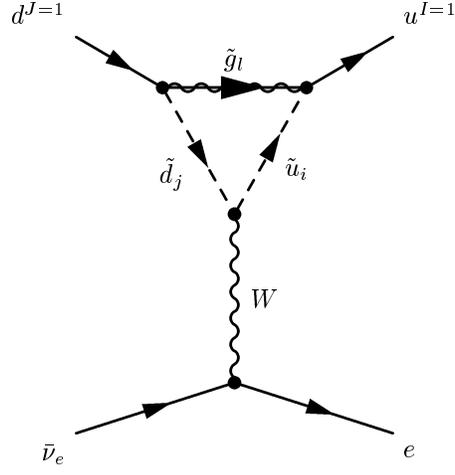}
 \caption{Vertex correction with gluino$-\tilde u$-$\tilde d$ loop.}
 \label{formeln:diag_gluino}
\end{center}
\end{figure}

\beq \label{eloop2}
\frac{\di \Gamma_2}{\di E_e \di\cos\theta_{e\bar\nu}} 
=& \frac{\di \Gamma_{\text{tree}}}{\di E_e \di\cos\theta_{e\bar\nu}}
    \cdot \vec{s_n} \cdot \left( 
    \frac{\vec{p}_e \times \vec{p}_{\bar\nu}}{E_e E_{\bar\nu}} \right) 
    \frac43 \frac1{4 m_n U^2 V^2} 
   \frac{ \Re{\left(UV^2\right)} } {16\pi^2} \nonumber \\ 
&\Bigg[ \Bigl( -2 \Im\left(A_2 D_2 E_2\right) m_p \Bigr)
\intkkd{u}{_i}{d}{_j}{g}{_l} \\
& + \Bigl( 4 \Im\left(A_2 C_2 E_2\right) 
  + 4 \Im\left(B_2 D_2 E_2\right) \Bigr) m_{\tilde{g}_l} m_n^2 
\intkdz{u}{_i}{d}{_j}{g}{_l} \nonumber \\
& + 4 \Im\left(B_2 C_2 E_2\right) m_n^3
\left( \intkdz{u}{_i}{d}{_j}{g}{_l} 
       - \intkdd{u}{_i}{d}{_j}{g}{_l} \right) 
\Bigg] \nonumber \\
& + \dots \nonumber
\eeq
with
\ben \label{eloopcoup2} \beq
A_2 &= g_3 \sqrt2 Z_U^{\left(I+3\right)i} \\
B_2 &= - g_3 \sqrt2 Z_U^{Ii} \\
C_2 &= - g_3 \sqrt2 Z_D^{Jj} \\
D_2 &= g_3 \sqrt2 Z_D^{\left(J+3\right)j} \\
E_2 &= Z_D^{Jj*} Z_U^{Ii*} C^{IJ} 
\eeq \een

\paragraph{Diagram 3: Neutralino--Chargino$-\tilde d$ vertex correction}~\\

\begin{figure}[ht]
\begin{center}
 \includegraphics{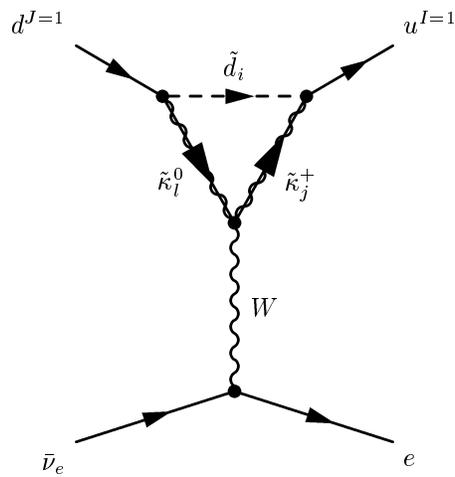}
 \caption{Vertex correction with neutralino--chargino$-\tilde d$ loop.}
\end{center}
\end{figure}

\beq \label{eloop3}
\frac{\di \Gamma_3}{\di E_e \di\cos\theta_{e\bar\nu}} =&
\frac{\di \Gamma_{\text{tree}}}{\di E_e \di\cos\theta_{e\bar\nu}}
    \cdot \vec{s_n} \cdot \left( 
    \frac{\vec{p}_e \times \vec{p}_{\bar\nu}}{E_e E_{\bar\nu}} \right) 
    \frac1{2m_n U^2 V^2} 
   \frac{ \Re{\left(UV^2\right)} } {16\pi^2} \nonumber \\ 
&\Bigg[ \Bigl( - \Im\left(A_3 C_3 F_3\right) m_p \Bigr) 
        \intkkd{\kappa}{^0_l}{\kappa}{^+_j}{d}{_i} \nonumber \\
& + 2 \Bigl( \Im\left(A_3 D_3 F_3\right) m_p m_{\tilde{\kappa}^+_j}
                m_{\tilde{\kappa}^0_l} 
      \intkdn{\kappa}{^0_l}{\kappa}{^+_j}{d}{_i} \nonumber \\
& \qquad + \Im\left(B_3 D_3 E_3\right) m_n^3 
      \intkdd{\kappa}{^0_l}{\kappa}{^+_j}{d}{_i} \\
& \qquad+ \left( \Im\left(A_3 D_3 E_3\right) m_{\tilde \kappa^+_j} 
+ \Im\left(B_3 D_3 F_3\right)
     m_{\tilde{\kappa}^0_l}  \right) m_n^2
    \intkdz{\kappa}{^0_l}{\kappa}{^+_j}{d}{_i}   \Bigr) \Bigg]
\nonumber \\ & + \dots \nonumber
\eeq
with
\ben \label{eloopcoup3} \beq
A_3 =& Y_u^I Z_D^{Mi*} Z_+^{2j} C^{IM} \\
B_3 =& - \left( \frac{e}{\sin\theta_W} Z_D^{Mi*} Z_-^{1j*} 
        + Y_d^I Z_D^{\left(M+3\right)i*} Z_-^{2j*} \right) C^{IM} \\
C_3 =& \frac{e}{\sin\theta_W} \left( Z_+^{1j*} Z_N^{2l} 
        - \frac1{\sqrt2} Z_+^{2j*} Z_N^{4l} \right) \\
D_3 =& \frac{e}{\sin\theta_W} \left( Z_-^{1j} Z_N^{2l*} 
        + \frac1{\sqrt2} Z_-^{2j} Z_N^{3l*} \right) \\
E_3 =& \left( - \frac{e}{\sqrt2\sin\theta_W\cos\theta_W} Z_D^{Ji}
        \left( \frac13
        Z_N^{1l} \sin\theta_W - Z_N^{2l} \cos\theta_W \right)
+ Y_d^J Z_D^{\left(J+3\right)i} Z_N^{3l} \right) \delta^{JM} \\
F_3 =& \left( \frac{-\sqrt2e}{3\cos\theta_W} Z_D^{\left(J+3\right)i} Z_N^{1l*}
        +Y_d^J Z_D^{Ji} Z_N^{3l*} \right) \delta^{JM}
\eeq \een

\paragraph{Diagram 4: Chargino-Neutralino-$\tilde u$ vertex correction}~\\

\begin{figure}[ht]
\begin{center}
 \includegraphics{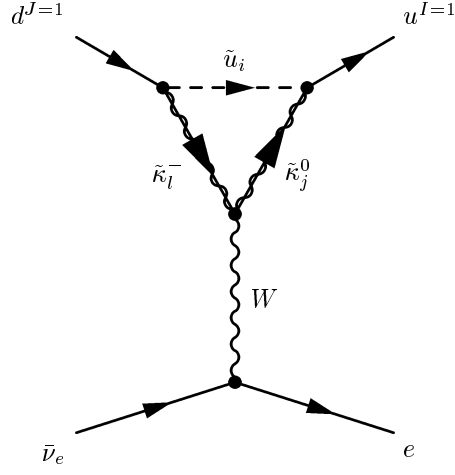}
 \caption{Vertex correction with chargino--neutralino-$\tilde u$ loop.}
\end{center}
\end{figure}

\beq \label{eloop4}
\frac{\di \Gamma_4}{\di E_e \di\cos\theta_{e\bar\nu}} 
=& \frac{\di \Gamma_{\text{tree}}}{\di E_e \di\cos\theta_{e\bar\nu}}
    \cdot \vec{s_n} \cdot \left( 
    \frac{\vec{p}_e \times \vec{p}_{\bar\nu}}{E_e E_{\bar\nu}} \right) 
    \frac1{2m_n U^2 V^2} 
   \frac{ \Re{\left(UV^2\right)} } {16\pi^2} \nonumber \\ 
&\Bigg[ \Bigl( - \Im\left(A_4 C_4 F_4\right) m_p \Bigr) 
  \intkkd{\kappa}{^-_l}{\kappa}{^0_j}{u}{_i} \nonumber \\
& + 2 \Bigl( \Im\left(A_4 D_4 F_4\right) m_p m_{\tilde{\kappa}^0_j}
                m_{\tilde{\kappa}^-_l}
     \intkdn{\kappa}{^-_l}{\kappa}{^0_j}{u}{_i} \nonumber \\
& \qquad + \Im\left(B_4 D_4 E_4\right) m_n^3 
     \intkdd{\kappa}{^-_l}{\kappa}{^0_j}{u}{_i} \\
& \qquad + \left( \Im\left(A_4 D_4 E_4\right) m_{\tilde \kappa^0_j}
+ \Im\left(B_4 D_4 F_4\right)
     m_{\tilde{\kappa}^-_l}  \right) m_n^2
 \intkdz{\kappa}{^-_l}{\kappa}{^0_j}{u}{_i} \Bigr) \Bigg] \nonumber \\
& + \dots \nonumber
\eeq
with
\ben \label{eloopcoup4} \beq
A_4 =& \left( \frac{2\sqrt2e}{3\cos\theta_W} Z_U^{\left(I+3\right)i} Z_N^{1j}
        -Y_u^I Z_U^{Ii} Z_N^{4j} \right) \delta^{IM} \\
B_4 =& - \left[ \frac{e}{\sqrt2\sin\theta_W\cos\theta_W} Z_U^{Ii} 
        \left( \frac13
        Z_N^{1j*} \sin\theta_W + Z_N^{2j*} \cos\theta_W \right)
        + Y_u^I Z_U^{\left(I+3\right)i} Z_N^{4j*} \right] \delta^{IM} \\
C_4 =& - \frac{e}{\sin\theta_W} \left( Z_-^{1l} Z_N^{2j*} 
        + \frac1{\sqrt2} Z_-^{2l} Z_N^{3j*} \right) \\
D_4 =& \frac{e}{\sin\theta_W} \left( - Z_+^{1l*} Z_N^{2j} 
        + \frac1{\sqrt2} Z_+^{2l*} Z_N^{4j} \right) \\
E_4 =& \left( - \frac{e}{\sin\theta_W} Z_U^{Mi*} Z_+^{1l} 
        + Y_u^M Z_U^{\left(M+3\right)i*} Z_+^{2l} \right) C^{MJ}\\
F_4 =& - Y_d^J Z_U^{Mi*} Z_-^{2l*} C^{MJ}
\eeq \een

\paragraph{Diagram 5: Box with $\tilde{u}$}~\\

\begin{figure}[ht]
\begin{center}
 \includegraphics{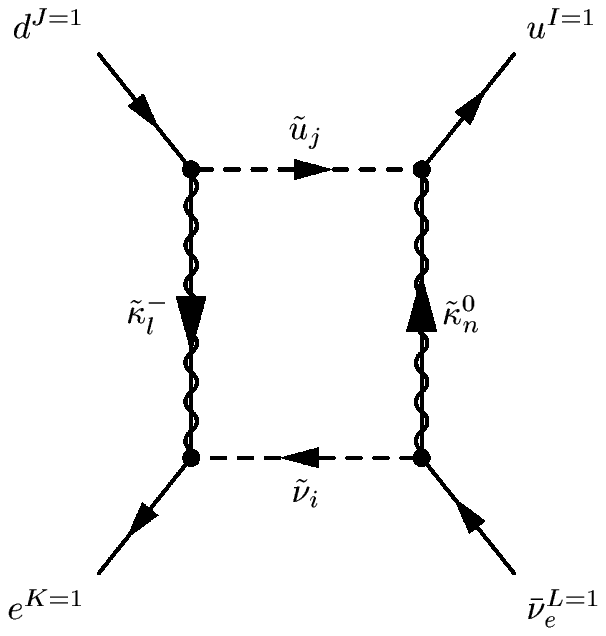}
 \caption{Box diagram with $\tilde{u}$ in the loop.}
\end{center}
\end{figure}

\beq \label{eloop5}
\frac{\di \Gamma_5}{\di E_e \di\cos\theta_{e\bar\nu}} =&
 \frac{\di \Gamma_{\text{tree}}}{\di E_e \di\cos\theta_{e\bar\nu}}
    \cdot \vec{s_n} \cdot \left( 
    \frac{\vec{p}_e \times \vec{p}_{\bar\nu}}{E_e E_{\bar\nu}} \right) 
    \frac{m_W^2}{4m_n U^2 V^2} 
   \frac{ \Re{\left(UV\right)} } {16\pi^2} \nonumber \\ 
&\Bigg\{ \Bigl( \Im\left(A_5 D_5 F_5 G_5 \right) m_e \Bigr) 
  \intkkv{u}{_j}{\nu}{_i}{\kappa}{^0_n}{\kappa}{^-_l} \nonumber \\
& + 2 \Bigl[ \Im\left(B_5 D_5 E_5 G_5\right) m_p m_{\tilde{\kappa}^-_l}
    m_{\tilde{\kappa}^0_n}
     \intkvn{u}{_j}{\nu}{_i}{\kappa}{^0_n}{\kappa}{^-_l} \nonumber \\
& \quad + \left( \Im\left(B_5 D_5 F_5 G_5\right) m_{\tilde{\kappa}^-_l} m_n^2 
 + \Im\left(B_5 C_5 E_5 G_5\right) m_{\tilde{\kappa}^0_n} m_n^2
\right) \nonumber \\
& \qquad  \intkve{u}{_j}{\nu}{_i}{\kappa}{^0_n}{\kappa}{^-_l} \\
& \quad + \Im\left(B_5 C_5 F_5 G_5\right) m_n^3 
     \intkvz{u}{_j}{\nu}{_i}{\kappa}{^0_n}{\kappa}{^-_l} \Bigr]
\Bigg\} \nonumber \\ & + \dots \nonumber
\eeq
with
\ben \label{eloopcoup5} \beq
A_5 =& - Y_l^K Z_\nu^{Ki} Z_-^{2l-} \\
B_5 =& - \frac{e}{\sin\theta_W} Z_\nu^{Ki} Z_+^{1l*} \\
C_5 =& - \left( \frac{e}{\sin\theta_W} Z_U^{Mj*} Z_+^{1l} 
        + Y_u^M Z_U^{\left(M+3\right)j*} Z_+^{2l} \right) C^{MJ} \\
D_5 =& - Y_d^J Z_U^{Mj*} Z_-^{2l*} C^{MJ} \\
E_5 =& \left( \frac{2\sqrt2e}{3\cos\theta_W}Z_U^{\left(I+3\right)j}Z_N^{1n}
        - Y_u^I Z_N^{4n} Z_U^{Ij} \right) \delta^{MI} \\
F_5 =& \left[ - \frac{e}{\sqrt2\sin\theta\cos\theta_W} Z_U^{Ij} \left( \frac13
        Z_N^{1n*} \sin\theta + Z_N^{2n*} \cos\theta\right)
- Y_u^I Z_N^{4n*}Z_U^{\left(I+3\right)j} \right] \delta^{MI} \\
G_5 =& \frac{e}{\sqrt2\sin\theta\cos\theta_W} Z_\nu^{Li*}
        \left( Z_N^{1n} \sin\theta - Z_N^{2n} \cos\theta \right)
\eeq \een

\paragraph{Diagram 6: Box with $\tilde{d}$}~\\

\begin{figure}[ht]
\begin{center}
 \includegraphics{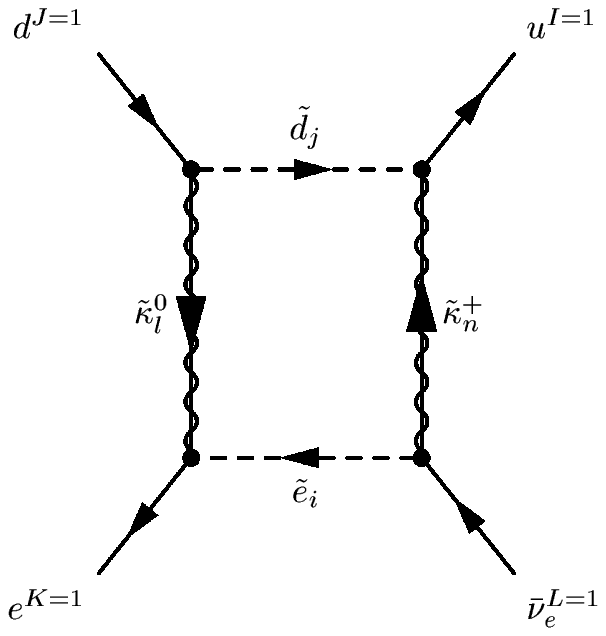}
 \caption{Box diagram with $\tilde{d}$ in the loop.}
\end{center}
\end{figure}

\beq \label{eloop6}
\frac{\di \Gamma_6}{\di E_e \di\cos\theta_{e\bar\nu}}
=& \frac{\di \Gamma_{\text{tree}}}{\di E_e \di\cos\theta_{e\bar\nu}}
    \cdot \vec{s_n} \cdot \left( 
    \frac{\vec{p}_e \times \vec{p}_{\bar\nu}}{E_e E_{\bar\nu}} \right) 
    \frac{m_W^2}{4m_n U^2 V^2} 
   \frac{ \Re{\left(UV\right)} } {16\pi^2} \nonumber \\ 
&\Bigg\{ \Bigl( \Im\left(A_6 D_6 F_6 G_6 \right) m_e \Bigr) 
  \intkkv{d}{_j}{e}{_i}{\kappa}{^+_n}{\kappa}{^0_l} \nonumber \\
& + 2 \Bigl[ \Im\left(B_6 D_6 E_6 G_6\right) m_p m_{\tilde{\kappa}^0_l}
    m_{\tilde{\kappa}^+_n}
     \intkvn{d}{_j}{e}{_i}{\kappa}{^+_n}{\kappa}{^0_l} \nonumber \\
& \quad + \left( \Im\left(B_6 D_6 F_6 G_6\right) m_{\tilde{\kappa}^0_l} m_n^2 
  + \Im\left(B_6 C_6 E_6 G_6\right) m_{\tilde{\kappa}^+_n} m_n^2
\right) \nonumber \\
& \qquad \intkve{d}{_j}{e}{_i}{\kappa}{^+_n}{\kappa}{^0_l} \\
& \quad + \Im\left(B_6 C_6 F_6 G_6\right) m_n^3 
    \intkvz{d}{_j}{e}{_i}{\kappa}{^+_n}{\kappa}{^0_l} \Bigr] \Bigg\}
\nonumber \\ & + \dots \nonumber
\eeq
with
\ben \label{eloopcoup6} \beq
A_6 =& - \frac{\sqrt2e}{\cos\theta_W}Z_L^{\left(K+3\right)i*}Z_N^{1l}
        + Y_l^K Z_N^{3l} Z_L^{Ki*} \\
B_6 =& \frac{e}{\sqrt2\sin\theta\cos\theta_W} Z_L^{Ki*} 
        \left( Z_N^{1l*} \sin\theta + Z_N^{2l*} \cos\theta \right)
        + Y_l^K Z_N^{3l*} Z_L^{\left(K+3\right)i*} \\
C_6 =& \left( \frac{-e}{\sqrt2\sin\theta\cos\theta_W} Z_D^{Jj}
        \left( \frac13 Z_N^{1l} \sin\theta - Z_N^{2l} \cos\theta \right) 
        + Y_d^J Z_D^{\left(J+3\right)j} Z_N^{3l} \right) \delta^{MJ} \\
D_6 =& \left( - \frac{\sqrt2e}{3\cos\theta_W}Z_D^{\left(J+3\right)j} Z_N^{1l*}
        + Y_d^J Z_N^{3l*} Z_D^{Jj} \right) \delta^{MJ} \\
E_6 =& - Y_u^M Z_D^{Ij*} Z_+^{2n} C^{IM} \\
F_6 =& - \left( \frac{e}{\sin\theta_W} Z_D^{Ij*} Z_-^{1n*} 
          + Y_d^I Z_D^{\left(I+3\right)j} Z_-^{2n*} \right) C^{IM} \\
G_6 =& - \left( \frac{e}{\sin\theta_W} Z_L^{Li} Z_-^{1n} 
        + Y_l^L Z_L^{\left(L+3\right)i} Z_-^{2n} \right)  
\eeq \een


%
%

\section{Loop Integrals}
\renewcommand{\theequation}{C.\arabic{equation}}
\setcounter{equation}{0}

In this Appendix we have summarized the definitions of the loop integrals
which were used in the calculation of the Feynman diagrams in the
previous Appendix. 

\subsection{2--point Functions}

First we define the 2--point functions as in \cite{Passarino:1979jh}
because most 3-- and 4--point functions will be expressed in terms of
these.

\begin{align}
 B_0(q^2,m_1,m_2) &= \Delta - \int_0^1 \di x\: \ln H \\
 B_1(q^2,m_1,m_2) &= -\frac12 \Delta + \int_0^1 \di x\: x \ln H \\
 B_{21}(q^2,m_1,m_2) &= \frac13 \Delta - \int_0^1 \di x\: x^2 \ln H \\
 B_3(q^2,m_1,m_2) &= - B_1(q^2,m_1,m_2) - B_{21}(q^2,m_1,m_2)
\nonumber\\
  &= \frac16 \Delta - \int_0^1 \di x\: x(1-x) \ln H \\
\intertext{with}
 H &= \left[(1-x)m_1^2 + x m_2^2 - x (1-x) q^2 -i\epsilon \right] \\
 \Delta &= \frac1\epsilon -\gamma_E + \ln 4\pi
\end{align}
where $\gamma_E = - \frac{\di \ln\Gamma(x)}{\di x}\Big|_{x=1} 
= 0,577216\dots$ denotes the Euler constant. Since the one--loop
corrections to $D$ are finite, the terms $\propto \Delta$ cancel in
the combinations of $B-$functions that will appear below.

\subsection{3--point Functions}

Besides the standard integrals $C_0$ and $C_{\mu\nu}$ with vanishing
external momenta, we need the following integrals which contain
combinations of the Feynman parameters in the nominator.
\begin{align}
C_{D1}&(0,0,m_1,m_2,m_3) = \int_0^1 \di x \int_0^{1-x} \di y 
   \frac{x+y}{m_1^2\left(1-x-y\right) + m_2^2 x + m_3^2 y} \nonumber\\
          &= \frac1{m_3^2 - m_2^2} 
 \left( B_0\left(0,m_2,m_1\right) - B_0\left(0,m_3,m_1\right) 
       - B_1\left(0,m_2,m_1\right) + B_1\left(0,m_3,m_1\right) \right) \\
C_{D2}&(0,0,m_1,m_2,m_3) = \int_0^1 \di x \int_0^{1-x} \di y 
   \frac{1-x-y}{m_1^2\left(1-x-y\right) + m_2^2 x + m_3^2 y} \nonumber\\
          &= \frac1{m_3^2 - m_2^2} 
 \left( B_1\left(0,m_2,m_1\right) - B_1\left(0,m_3,m_1\right) \right) \\
C_{D3}&(0,0,m_1,m_2,m_3) = \int_0^1 \di x \int_0^{1-x} \di y 
   \frac{\left(1-x-y\right)^2}{m_1^2\left(1-x-y\right) + m_2^2 x + m_3^2 y} 
      \nonumber\\
          &= \frac1{m_3^2 - m_2^2} 
 \left( B_1\left(0,m_2,m_1\right) - B_1\left(0,m_3,m_1\right) 
       - B_3\left(0,m_2,m_1\right) + B_3\left(0,m_3,m_1\right) \right) 
\end{align}

\subsection{4--point Functions}

Here the standard integrals $D_0$ and $D_{\mu\nu}$ with vanishing
external momenta are needed. In addition the following two integrals
with Feynman parameters appear:
\begin{align}
D_{D1}&(0,0,0,m_1,m_2,m_3,m_4) \nonumber\\
 &=\int_0^1 \di x \int_0^{1-x} \di y \int_0^{1-x-y} \di z
   \frac{x}{ \left[ m_1^2\left(1-x-y-z\right) 
     + m_2^2 x + m_3^2 y + m_4^2 z \right]^2 } \nonumber\\
 &= \frac1{m_4^2 - m_1^2}
     \Big( \frac1{m_3^2 - m_4^2}
      \left( B_1\left(0,m_3,m_2\right) - B_1\left(0,m_4,m_2\right) \right)
       \nonumber\\
           & \hspace{60pt} - \frac1{m_3^2 - m_1^2}
      \left( B_1\left(0,m_3,m_2\right) - B_1\left(0,m_1,m_2\right) \right)
      \Big) \\
D_{D2}&(0,0,0,m_1,m_2,m_3,m_4) \nonumber\\
 &=\int_0^1 \di x \int_0^{1-x} \di y \int_0^{1-x-y} \di z
   \frac{x^2} { \left[ m_1^2\left(1-x-y-z\right) 
     + m_2^2 x + m_3^2 y + m_4^2 z \right]^2 } \nonumber\\
 &= \frac1{m_4^2 - m_1^2}
     \Big( \frac1{m_3^2 - m_4^2}
      \big( B_1\left(0,m_3,m_2\right) - B_1\left(0,m_4,m_2\right)
             \nonumber\\
 &\hspace{120pt} - B_3\left(0,m_3,m_2\right) + B_3\left(0,m_4,m_2\right)\big)
       \nonumber\\
           & \hspace{60pt} - \frac1{m_3^2 - m_1^2}
      \big( B_1\left(0,m_3,m_2\right) - B_1\left(0,m_1,m_2\right) 
            \nonumber\\
 &\hspace{120pt} - B_3\left(0,m_3,m_2\right) + B_3\left(0,m_1,m_2\right)\big) 
      \Big)
\end{align}

\end{appendix}


\end{document}